\documentclass[twocolumn]{aastex63}

\usepackage{xcolor}

\definecolor{myred}{HTML}{BF3465}

\def\ltsima{$\; \buildrel < \over \sim \;$}
\def\simlt{\lower.5ex\hbox{\ltsima}}
\def\gtsima{$\; \buildrel > \over \sim \;$}
\def\simgt{\lower.5ex\hbox{\gtsima}}

\newcommand {\um}{$\mu$m}

\newcommand {\msun}{M$_{\odot}$}

\newcommand {\Cii}{[C\textsc{ii}]}
\newcommand {\Oiii}{[O\textsc{iii}]}

\newcommand {\pz}{$P(z)$}

\newcommand {\hband}{$H_{160}$}
\newcommand {\jband}{$J_{125}$}
\newcommand {\yband}{$Y_{105}$}
\newcommand {\iband}{$I_{814}$}
\newcommand {\vband}{$V_{606}$}
\newcommand {\jhband}{\textit{JH}$_{140}$}
\newcommand {\yjband}{\textit{YJ}$_{110}$}
\newcommand {\zband}{$z_{850}$}

\def\ltsima{$\; \buildrel < \over \sim \;$}
\def\simlt{\lower.5ex\hbox{\ltsima}}
\def\gtsima{$\; \buildrel > \over \sim \;$}
\def\simgt{\lower.5ex\hbox{\gtsima}}

\received{--}
\revised{--}
\accepted{30 May 2023}
\submitjournal{ApJ}

\shorttitle{LBG search in the fields of $z\sim6$ quasars}
\shortauthors{Champagne et al.}

\begin{document}

\title{A Mixture of LBG Overdensities in the Fields of Three $6<z<7$ Quasars: Implications for the Robustness of Photometric Selection}

\correspondingauthor{Jaclyn B. Champagne}
\email{jbchampagne@arizona.edu}

\author[0000-0002-6184-9097]{Jaclyn~B.~Champagne}
\affiliation{Department of Astronomy, University of Texas at Austin, Austin, TX 78712}
\affiliation{Steward Observatory, University of Arizona, 933 N. Cherry Ave, Tucson, AZ 85721}
\author[0000-0002-0930-6466]{Caitlin~M.~Casey}
\affiliation{Department of Astronomy, University of Texas at Austin, 2515 Speedway Blvd, Austin, TX 78712}
\affiliation{Cosmic Dawn Center (DAWN), Denmark}
\author[0000-0001-8519-1130]{Steven~L.~Finkelstein}
\affiliation{Department of Astronomy, University of Texas at Austin, 2515 Speedway Blvd, Austin, TX 78712}
\author[0000-0002-9921-9218]{Micaela~Bagley}
\affiliation{Department of Astronomy, University of Texas at Austin, 2515 Speedway Blvd, Austin, TX 78712}
\author[0000-0003-3881-1397]{Olivia~R.~Cooper}
\affiliation{Department of Astronomy, University of Texas at Austin, 2515 Speedway Blvd, Austin, TX 78712}
\author[0000-0003-2366-8858]{Rebecca~L.~Larson}
\affiliation{Department of Astronomy, University of Texas at Austin, 2515 Speedway Blvd, Austin, TX 78712}
\author[0000-0002-7530-8857]{Arianna~S.~Long}
\affiliation{Department of Astronomy, University of Texas at Austin, 2515 Speedway Blvd, Austin, TX 78712}
\author[0000-0002-7633-431X]{Feige Wang}
\affiliation{Steward Observatory, University of Arizona, 933 N. Cherry Ave, Tucson, AZ 85721}

\begin{abstract}
The most luminous quasars at $z>6$ are suspected to be both highly clustered and reside in the most massive dark matter halos in the early Universe, making them prime targets to search for galaxy overdensities and/or protoclusters.
We search for Lyman-break dropout-selected galaxies using \textit{HST} WFC3/ACS broadband imaging in the fields of three $6<z<7$ quasars, as well as their simultaneously observed coordinated-parallel fields, and constrain their photometric redshifts using EAZY.
One field, J0305-3150, shows a volume density 10$\times$ higher than the blank-field UV luminosity function (UVLF) at $M_{UV} < -20$, with tentative evidence of a 3$\sigma$ overdensity in its parallel field located 15 cMpc away.
Another field, J2054-0005, shows an angular overdensity within 500 ckpc from the quasar but still consistent with UVLF predictions within 3$\sigma$, while the last field, J2348-3054, shows no enhancement. 
We discuss methods for reducing uncertainty in overdensity measurements when using photometric selection and show that we can robustly select LBGs consistent with being physically associated with the quasar, corroborated by existing \textit{JWST}/NIRCam WFSS data in the J0305 field.
Even accounting for incompleteness, the overdensities in J0305 and J2054 are higher for brighter galaxies at short angular separations, suggesting preferential enhancement of more massive galaxies in the quasar's immediate vicinity. 
Finally, we compare the LBG population with previously-identified \Cii\, and mm-continuum companions; the LBG overdensities are not accompanied by an enhanced number of dusty galaxies, suggesting that the overdense quasar fields are not in the bursty star-forming phase sometimes seen in high-redshift protoclusters.

\end{abstract}

\keywords{galaxies, quasars}

\section{Introduction} \label{sec:intro}
Numerous ground-based surveys undertaken within the last ten years \citep[e.g.,][]{Venemans2013a, Banados2016a, Wang2019a, Yang2019a, Wang2021a} have revealed a population of $\gtrsim60$ rare, bright quasars lying at $z>6.5$ with black hole masses $\gtrsim10^9\, \rm M_{\odot}$ \citep[e.g.,][]{Banados2018a, Yang2021a, Farina2022a}. 
The very existence of quasars powered by such massive black holes $<1$ Gyr after the Big Bang challenges our understanding of supermassive black hole (SMBH) formation.
A number of evolutionary scenarios have been tested in cosmological models to produce the observed spatial density of these extraordinary quasars, including super-Eddington accretion %\citep{Inayoshi2016a} 
or $\rm >1000\, M_{\odot}$ seed black holes \citep[see review in][]{Volonteri2021a}. 
Observations suggest SMBHs are hosted by massive ($M_* > 10^{11}\, \rm M_{\odot}$) galaxies with prodigious star formation rates (SFR)  $> 10^2 \,\rm M_{\odot}\, yr^{-1}$  \citep[e.g.,][]{Decarli2018a}.
These objects are expected to form in massive halos that collapsed early or experienced large inflow rates --- suggestive of residing in overdense regions of the cosmic web --- though some simulations predict large variance in the number of galaxies expected around quasars \citep[e.g.,][]{Overzier2009a, Habouzit2019a}.

In the $\Lambda$CDM paradigm, structures form hierarchically \citep{Vogelsberger2014a}, from primordial density perturbations to the virial collapse of the galaxy clusters after $z\sim1.5$.
Cosmic downsizing suggests that the most massive objects assembled their mass more quickly than objects in lower density regions \citep{Cowie1986a}, implying that the seeds of $z=0$ galaxy clusters should exist in massive overdensities at very high redshifts.
Thus, it is natural to conclude that quasars and galaxy overdensities trace the same large-scale dark matter density fluctuations at early times.
This is corroborated by observations and models at least out to $z\sim4$ that find a strong angular auto-correlation between quasars \citep{Shen2007a} and cross-correlation between quasars and galaxies through both individual detections of Lyman-break galaxies (LBGs) and \textit{Herschel} stacks tracing dusty star-forming galaxies (DSFGs) \citep{Garcia-Vergara2017a, Hall2018a}. These imply that some quasars live in highly clustered environments at $z\sim4$, yet positive quasar-galaxy cross-correlation results are also not ubiquitous \citep[e.g.,][]{Adams2015a, Uchiyama2018a, Uchiyama2020a}.

High-redshift protoclusters or overdensities have thus typically been identified via number counts of star-forming galaxies in the optical/infrared (OIR) and millimeter, for example with Lyman-alpha emitters (LAEs), LBGs, or DSFGs, many of them centered on massive central objects like radio galaxies \citep[e.g.,][]{Venemans2007a, Intema2006a, Walter2012a, Hennawi2015a}. 
But the bias by galaxy type and redshift --- i.e., how strongly a class of galaxies traces the underlying dark matter overdensity --- is relatively unknown. 
If galaxy protoclusters primarily build their mass through intense, stochastic starburst episodes, their overdensities may manifest as heavily dust-obscured galaxies \citep{Casey2016a}. 
In fact, obscured galaxies might be expected in higher excesses than LAEs or LBGs around quasars if ionizing UV radiation from the quasar suppresses lower-mass galaxy formation in its immediate vicinity \citep[i.e., the proximity zone;][]{Satyavolu2022a}.
On the other hand, if DSFGs are not ubiquitous features of protocluster populations at any given time --- which may indeed be the case since they are rare and short-lived \citep*[see review in][]{Casey2014a} --- we might instead find overdensities via the less massive, modestly star-forming population traced by LBGs.

Yet evidence for overdensities around quasars at $z>5$ remains elusive in both the rest-frame UV and far infrared.
For example, $4-10\sigma$ enhancements of LAEs/LBGs \citep{Kashikawa2007a, Utsumi2010a, Mignoli2020a} and \Cii\, emitters \citep{Decarli2017a} have been observed around $z>6$ quasars, while others find no excesses in dust continuum \citep{Champagne2018a, Meyer2022a} or LAEs/LBGs \citep{Banados2013a, Mazzucchelli2017a, Uchiyama2017a}. 
Still others find a mix of over- and underdensities within the same samples \citep{Kim2009a, Ota2018a}.
Much of the uncertainty in these results comes from a lack of sensitivity and/or spectroscopic completeness, i.e. the volume occupied by these galaxy populations is unconstrained due to varying selection criteria within different coverage areas.
%[more references in Garcia-Vergara]
Searches for UV-bright galaxies are able to be performed in fields extending tens of arcminutes, more comparable to the expected sizes of protoclusters \citep[$>0.5$\,deg on the sky by $z=2$, e.g.,][]{Muldrew2015a, Hung2016a}, but due to the small fields of view in single-pointing millimeter observations, it is not possible to map one-to-one the varying spatial distributions of obscured versus unobscured galaxies without wide mosaics.
Nonetheless, expanding the sample of reionization-era quasars with environment studies is necessary to measure the statistical variation in their clustering strengths.

Finally, we do not yet know which types of galaxies accurately trace the underlying dark matter overdensities, particularly during the epoch of reionization.
The UV luminosity function is relatively uncertain during the EoR, particularly at the bright end \citep[e.g.,][]{Bowler2017a,Bagley2022a}, and completely unconstrained in clustered environments.
\textit{JWST} and the \textit{Roman Space Telescope} will provide the best opportunity to measure the clustering of massive galaxies in wide deep fields at $z>9$, but \textit{HST} remains powerful for selecting LBGs at $z\sim6$, allowing us to compare quasar fields with the blank field UVLF \citep[e.g.,][]{Bouwens2015a, Finkelstein2015a}.

\begin{longtable*}{lccccccc}

\caption{Observational properties of the quasars in this sample. References: (1) \citet{Venemans2017a}, (2) \citet{Jiang2008a}, (3) \citet{Venemans2013a}, (4) \citet{Farina2022a}.}\\
\hline\hline

\multicolumn{7}{c}{ }\\
Target name & RA & Dec & $z$ & $M_{1450}$ & M$_{BH}$  & References \\
 & (J2000) & (J2000) & & (AB mag) & (10$^9$ \msun)  & \\
\hline
\endfirsthead

\hline
\endfoot
    
    VIK J0305--3150 & 03:05:16.92 & $-$31:50:55.90 & 6.60 & $-$25.96 $\pm$ 0.06 & 0.77$^{+0.16}_{-0.18}$ & 1, 4\\ %& 44.0 & 6.6145 $\pm$ 0.0001 & \Cii\, & 15\\%VENEMANS17\\
    
    SDSS J2054--0005 & 20:54:06.49 & $-$00:05:14.80 & 6.04 & $-$26.11 $\pm$ 0.09 & 2.17$^{+0.27}_{-0.25}$ & 2, 4\\ %& 30.7 & 6.0391 $\pm$ 0.0022 & CO & 13\\%WANG13\\
    
    VIK J2348--3054 & 23:48:33.34 & $-$30:54:10.24 & 6.90 & $-$25.72 $\pm$ 0.14 & 3.25$^{+1.17}_{-0.93}$ & 3, 4 \\ %& 51.9 & 6.9018 $\pm$ 0.0007 & \Cii\, & 15\\%VENEMANS13\

%\enddata

\label{table:qsos}

\end{longtable*}

In this paper, we present \textit{HST} followup of three $z>6$ quasars selected from previous ALMA observations \citep{Champagne2018a, Decarli2018a}.
We search for LBGs within $\Delta z \approx 1$ using photometric redshift fitting and attempt to measure the angular scale of their overdensities.
This paper is structured as follows: in $\S$\ref{sec:obs} we describe the \textit{HST} and \textit{Spitzer} observations, data reduction, and noise characterization. 
$\S$\ref{sec:catalog} describes the detection parameters used to construct the catalogs. 
$\S$\ref{sec:eazy} describes our selection criteria and photometric redshift fitting procedure.
In $\S$\ref{sec:discussion2d} we present a 2D analysis of the clustering strength of the LBGs and we present the results of the 3D luminosity function in $\S$\ref{sec:3d}. 
Finally, $\S$\ref{sec:discussion} contains a discussion of the implications for quasar environments and $\S$\ref{sec:conclusion} concludes. 
We report all magnitudes in AB and assume the \textit{Planck} cosmology with a flat Universe, with $\rm H_0$ = 70 $\rm km\, s^{-1} Mpc^{-1}$ and $\Omega_m$ = 0.3 \citep{Planck2016a}.

\section{Observations and Data Reduction}\label{sec:obs}

In this study we focus on the environments of three luminous quasars which contained one or more nearby dust continuum or \Cii\, emitters in previous ALMA surveys \citep{Champagne2018a, Venemans2020a, Meyer2022a}.
In Table \ref{table:qsos} we list the observational properties of the quasars in this sample including black hole masses and UV magnitudes. 
We combine \textit{HST} broadband imaging with archival \textit{Spitzer}/IRAC data from Channels 1 and 2 and outline our data processing techniques below.

\subsection{HST and Spitzer Data}
\begin{figure}
    \centering
    \includegraphics[trim={0.05in 0.05in 0.05in 0.05in}, width=1.0\columnwidth]{ 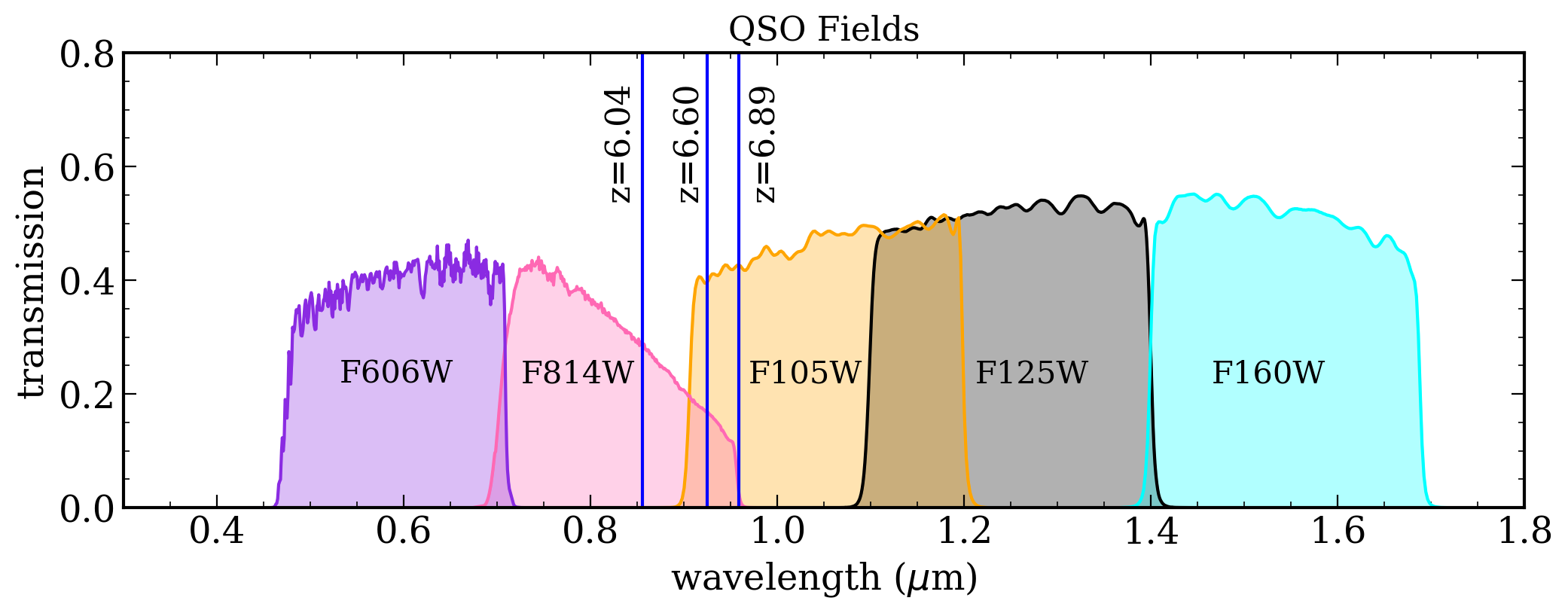}
    {\includegraphics[trim={0.05in 0.05in 0.05in 0.05in}, width=1.0\columnwidth]{  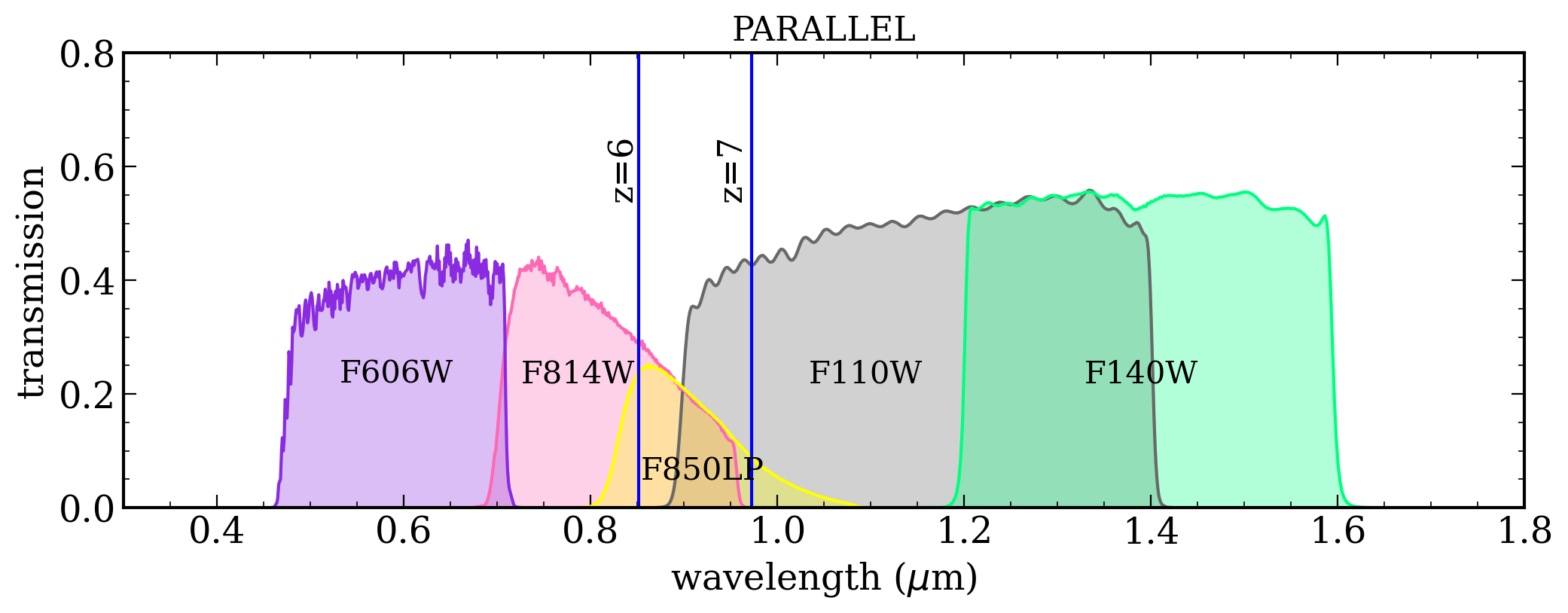}}
    \caption{Transmission curves for the quasar \textit{(top)} and parallel \textit{(bottom)} fields. The blue lines indicate the quasar redshifts, where LBG candidates are expected to drop out blueward of \yband.}
    \label{fig:transmission}
\end{figure}

The primary quasar fields were observed with ACS and WFC3 broadband filters between April 2018 and December 2019 (GO program \#15064, PI Casey), chosen to maximize the sensitivity to faint sources and allow for accurate comparisons to previous UVLF studies with \textit{HST}.
Each quasar was observed for one orbit each with ACS F606W (\vband) and F814W (\iband) and WFC3 F105W (\yband), F125W (\jband), and F160W (\hband), bracketing the Lyman break at observed-frame 0.85--1\,\um.
At these redshifts, the Lyman break occurs at rest-frame 1216\,\AA\, (Ly$\alpha$) due to the opacity of the IGM at $z>6$.
ACS and WFC3 were kept at a fixed orientation so that we could observe a coordinated parallel field with a similar filter set, under the hypothesis that extended protocluster structures would span well beyond the field of view of \textit{HST} and thus potentially yield an overdensity in a physically separated parallel field. 
The parallels were thus observed with ACS \vband, \iband, and F850LP (\zband), plus WFC3 F110W (\yjband) and F140W (\jhband).
Since J2054--0005 had existing F125W and F160W imaging (GO program \#12974, PI Mechtley), our coordinated parallel field has only \iband, \yjband\, and \jhband\, imaging in this program.
In Figure \ref{fig:transmission} we show the transmission curves for the quasar and parallel fields and where the redshifted Lyman break would fall within the \iband\, and \yband\, filters.

We also use archival \textit{Spitzer} IRAC imaging to aid in photometric redshift fitting, since the observed NIR colors of a Lyman break at $z=6$ can be degenerate with a Balmer break at $z\sim1-2$. 
All three quasars were observed with IRAC 3.6\,\um\, and 4.5\,\um\, (ID: 62291, PI Decarli).
These observations were intended to detect the quasar host galaxy; thus, they are relatively shallow for our purposes (reaching depths of 21.7 and 22.0 mag respectively), and there is no \textit{Spitzer} coverage of the parallel fields. 
In practice, we did not find that these images were deep enough to detect counterparts to any of the $z\sim6$ LBG candidates.
However, the inclusion of IRAC data allows us to easily discard lower redshift interlopers in photo-$z$ fitting, as faint \textit{HST} sources with bright IRAC counterparts are more likely to be line emitters at lower redshifts \citep[e.g.,][]{Finkelstein2022a}\footnote{Galaxies at $z=6-7$ can manifest as faint \textit{HST} detections with bright IRAC counterparts due to the presence of strong nebular emission (i.e.  H$\beta$+\Oiii) but their IRAC photometry is typically 2--3 mag fainter than the limit in this archival data \citep[see][]{Endsley2021a}.}.

\subsection{Point Spread Function Matching}\label{sec:psf}
We use the drizzled products delivered by the MAST archive and follow the photometric analysis techniques of \citet{Finkelstein2015a}, who used LBGs to measure the UV luminosity function at $z=6$ and $z=7$.
Because both the point spread function (PSF) and pixel scale are substantially different between ACS and WFC3 images, we must resample and PSF-match the images in order to correctly register the astrometry and measure accurate photometric colors.
We do this first by constructing empirical point spread functions for each field, then constructing a match kernel that is convolved with each image, such that the final image has the same spatial resolution as the worst PSF (0.13\arcsec\, in \hband\, and 0.15\arcsec\, in \jhband\, for primary and parallel fields respectively). 

We constructed empirical PSFs by stacking stars in every filter for each field. 
For all source detection, we use Source Extractor \citep[SE, version 2.19.5;][]{Bertin1996a}.
The initial catalog of stars was constructed with preliminary run of SE using the default Kron parameters ($k$=2.5, $R_{min}$=3.5), with candidate stars chosen to be in the stellar locus within the plane of half-light radius and magnitude (i.e., $m_{AB} < 22$ and radius $>3$ pixels).
The candidates were cross-matched with the \textit{Gaia} DR2 catalog \citep{Gaia2020a}, discarding anything with SE-generated flags indicating blended sources, saturation, or edge pixels. 
We created 101$\times$101-pixel cutouts of each star and stacked the normalized, background-subtracted images. 
We then manually set any remaining $>3\sigma$ pixels to the local rms value in the stacked image, subtracted any remaining background flux from the stacked image, and normalized the final PSF to unity. 

With empirical PSFs in hand, we created a matching convolution kernel using \verb|astropy|'s match-kernel function. 
This required a custom window function which we manually iterated upon until the light profile of stars in PSF-matched images matched that of the \hband\, or \jhband\, band images. 
Since we constructed the PSFs from the drizzled WFC3 and ACS images with their respective native pixel scales, the match kernel was resampled accordingly. 
Each image was convolved with the match kernel to match the spatial resolution of the \hband/\jhband\, images.

Finally, each PSF-matched image was projected to the same WCS header and pixel scale after aligning them to \textit{Gaia} DR2. 
We did this using astropy's \verb|reproject| function with exact interpolation and a final output pixel scale of 0.03\arcsec/pixel, reprojected to the WCS header of the ACS image. 
The reprojected, resampled images are flux-conserved and we took care to retroactively adjust the PSF-matching window function such that the curve-of-growth of stars in the final images was within 4\% of that in the \hband/\jhband\, images out to a default aperture diameter of 0.4\arcsec.

\begin{deluxetable*}{ccccccccc}
\tablehead{\colhead{field} & \colhead{F606W} & \colhead{F814W} & \colhead{F850LP} & \colhead{F105W} & \colhead{F110W} & \colhead{F125W} & \colhead{F140W} & \colhead{F160W}}
\tablecaption{Zeropoints for each field and filter, adjusted individually using calibration stars and their \textit{Gaia} and 2MASS photometry. $^{\dagger}$Archival data from GO program \#12974. \label{table:zpt}}
\startdata
J0305 QSO & 26.36 & 26.23 & $\cdots$ & 25.85 & $\cdots$ & 26.04 & $\cdots$ & 25.83 \\
J0305 PAR & 26.64 & 26.26 & 25.11 & $\cdots$ & 26.41 & $\cdots$ & 26.35 & $\cdots$ \\
J2054 QSO & 26.63 & 26.23 & $\cdots$ & 25.89 & $\cdots$ & 26.07$^{\dagger}$ & $\cdots$ & 25.81$^{\dagger}$ \\
J2054 PAR & $\cdots$ & 26.09 & $\cdots$ & $\cdots$ & 26.68 & $\cdots$ & 26.43 & $\cdots$ \\
J2348 QSO & 26.61 & 26.13 & $\cdots$ & 25.84 & $\cdots$ & 26.03 & $\cdots$ & 25.82 \\
J2348 PAR & 26.77 & 26.15 & 25.09 & $\cdots$ & 26.74 & $\cdots$ & 26.46 & $\cdots$ \\
\enddata
\end{deluxetable*}

\begin{deluxetable*}{ccccccccc}
\tablehead{\colhead{field} & \colhead{F606W} & \colhead{F814W} & \colhead{F850LP} & \colhead{F105W} & \colhead{F110W} & \colhead{F125W} & \colhead{F140W} & \colhead{F160W}}
\tablecaption{Limiting magnitudes at 5$\sigma$ for each field and filter, measured as the standard deviation in 0.4\arcsec\, apertures. $^{\dagger}$Archival data from GO program \#12974. \label{table:limmag}}
\startdata
J0305 QSO & 27.1 & 26.7 & $\cdots$ & 26.6 & $\cdots$ & 27.0 & $\cdots$ & 26.8 \\
J0305 PAR & 27.2 & 26.9 & 25.9 & $\cdots$ & 26.9 & $\cdots$ & 27.0 & $\cdots$ \\
J2054 QSO & 27.0 & 26.6 & $\cdots$ & 26.2 & $\cdots$ & 27.2$^{\dagger}$ & $\cdots$ & 26.8$^{\dagger}$ \\
J2054 PAR & $\cdots$ & 26.3 & $\cdots$ & $\cdots$ & 27.4 & $\cdots$ & 26.5 & $\cdots$ \\
J2348 QSO & 27.2 & 26.6 & $\cdots$ & 26.3 & $\cdots$ & 26.7 & $\cdots$ & 26.5 \\
J2348 PAR & 27.3 & 26.7 & 25.6 & $\cdots$ & 27.2 & $\cdots$ & 27.3 & $\cdots$
\enddata
\end{deluxetable*}

\subsection{Estimating Noise, Zeropoints, and Limiting Magnitudes}\label{sec:zpt}

In preliminary source catalogs, we noticed substantial systematic offsets between the expected colors of dropout sources and what we measured using the theoretical ACS and WFC3 zeropoints.
To check if the header zeropoints were accurate, we compared the \textit{HST} photometry of sources identified in the \textit{Gaia} DR2 star catalog \citep{Gaia2020a} and the 2MASS point source catalog \citep{Cutri2003a}.
Using the \citet{Pickles1998a} stellar spectral template library, we generated mock photometry for a subset of main sequence G, K, and M stars in the \textit{Gaia}, 2MASS, and \textit{HST} passbands (assuming a power-law extrapolation for the Rayleigh-Jeans tail of the stellar spectra) and constructed an empirical color relation as functions of both \textit{Gaia} BP$-$RP color and 2MASS $J-H$ color.
The ACS zeropoints in each filter were then adjusted to the median offset between \textit{Gaia} photometry and measured HST photometry for each filter, with a median offset of 0.15 mag, 0.25 mag, and 0.26 mag in \vband, \iband, and \zband\, respectively.
For WFC3, the zeropoints were adjusted via the offset between 2MASS photometry and \textit{HST}, but there were no 2MASS sources in the J0305 or J2348 quasar fields, so the offset from the J2054 quasar field photometry was applied to all fields.
These offsets were $-$0.36 for \yband, $-$0.16 for \jband, and $-$0.13 in \hband. 
Finally, the median offsets between 2MASS and \textit{HST} in the parallel fields were $-$0.12 for \yjband\, and $-$0.01 for \jhband.  
The adjusted zeropoints are listed in Table \ref{table:zpt}.
We find that these offsets remain the same even when considering the original, non-PSF-homogenized images, confirming that this offset is real and not due to our reprojection algorithm.
In Appendix \ref{sec:zptunc} we discuss the effects of our zeropoint assumptions on our measured photometry, noting that some caution is needed to interpret the absolute photometry but our SED results remain the same.

To estimate the uncertainty in our catalog fluxes, we constructed an empirical noise function as a function of aperture size since the noise is dependent on the number of pixels in the aperture. 
We measured the flux at 1000 positions distributed across each image (rejecting any apertures that contain edges, bad pixels, or real sources based on a preliminary segmentation map) with aperture sizes ranging from 1 to 20 pixels (0.6\arcsec). 
The limiting depth is then 5$\sigma$ where $\sigma$ is the standard deviation of fluxes measured in the 0.4\arcsec\, aperture and corrected to total from the curve-of-growth; these are listed in Table \ref{table:limmag}.
For individual sources, the SNR is determined as the interpolated value of the noise function given the area of the extraction aperture, multiplied by the ratio between the RMS value at the pixel centroid (a measure of the local background) and the overall RMS (a measure of the global background), based on the provided weight maps where RMS = $(\rm weight\, map)^{-1/2}$.

\section{Catalog Construction}\label{sec:catalog}

\subsection{Detection Parameters}\label{sec:params}

We ran Source Extractor (SE) in dual-image mode using \hband\, (\jhband) as the detection image for primary (parallel) fields with respective weight maps for each filter.
The relevant detection parameters are the following: \verb|DETECT_THRESH| = 1.0, Kron parameters \verb|k, Rmin| = 1.2, 1.7, \verb|DETECT_MINAREA| = 28 pixels, and \verb|PHOT_APERTURES| = 10 pixels.
These Kron parameters were chosen to maximize the sensitivity to faint unresolved sources, but we ran SE an additional time with the default Kron parameters (\verb|k, Rmin| = 2.5, 3.5) in order to derive an aperture correction for each source, defined as the ratio between the custom and default Kron fluxes in \hband\, and applied to the photometry in every filter.
We calculated the Galactic extinction in each filter by querying the \citet{Schlegel1998a} dust reddening map at the central position of each image and converting to $A_V$ using Table 6 of \citet{Schlafly2011a}.
To derive the final photometry, we applied the zeropoints, aperture corrections, and extinction correction to the small Kron auto flux, \verb|FLUX_AUTO|. 
%To derive photometry, we used \verb|FLUX_APER| to which we apply the aperture and extinction correction. 

We also incorporate the IRAC images but we do not PSF-match them to the WFC3 PSF due to the substantially coarser resolution; as such, we cannot use \textit{HST} as the detection image for these catalogs. 
Instead, the final \textit{HST} SE catalogs are matched by position to an IRAC catalog from a separate SE run and visually inspected.
Due to the shallow depth of the existing IRAC data, none of the $6<z<7$ candidates have individual point source counterparts, so we extract 3$\sigma$ upper limits in a 2\arcsec\, aperture with a warm-mission aperture correction of 6\% as prescribed by the IRAC Instrument Handbook.
Further, none of the candidates are contaminated by extended emission from nearby sources that are easily resolved by the \textit{HST} data, therefore no deblending was required.

\section{Searching for Lyman-break Galaxies}\label{sec:eazy}
\subsection{Fitting Photometric Redshifts with EAZY}
Historically, searches for LBGs have relied on two-color selection criteria that bracket the Lyman (or Ly$\alpha$) break and the rest-UV continuum, which evolve with redshift.
But, at $z>6$, the color criteria most notably suffer from contamination of low-luminosity stars whose colors are degenerate with blue ($\beta<0$) LBGs. 
In contrast, SED fitting has the advantage of using all of the available photometry simultaneously, although it is sensitive to the supplied templates, which carry more uncertainty for EoR galaxies that have lower metallicity, younger stellar populations, and bluer stellar continua \citep[e.g.,][]{Larson2022a}.
To obtain photometric redshifts, we use the EAZY \citep{Brammer2008a} software which fits a non-negative linear combination of template SEDs with the \textit{HST} and \textit{Spitzer} photometry. 
We use the newest FSPS default templates ({\tt tweak\_fsps\_QSF\_12\_v3}) in addition to a custom set of 12 templates optimized for the selection of young, blue LBGs at high redshift, which are described in detail in \citet[][hereafter L22]{Larson2022a}. 

Briefly, the L22 templates are generated from BPASS \citep{Eldridge2017a} and \verb|Cloudy| \citep{Ferland1998a} with a range of stellar population ages and line emission strengths. 
Three BPASS models include binary stars, a Chabrier IMF with an upper mass limit of 100 $\rm M_{\odot}$, a metallicity of $Z=0.05 Z_{\odot}$, and stellar ages of log(age/yr) = 6, 6.5, and 7.
A second version of each template also contains Ly$\alpha$ and high-EW nebular optical emission lines from \verb|Cloudy|. %(which in practice do not have a strong effect on our fits since our photometry is blueward of observed-frame 1.6\um).
The final two variations reduce the Ly$\alpha$ line strength to 1/3 of that produced by \verb|Cloudy| and remove the line entirely, which is a reasonable physical scenario at these redshifts where Ly$\alpha$ may still be attenuated by a significantly neutral IGM.
We eliminate the full-strength Ly$\alpha$ templates as the visibility of Ly$\alpha$ is expected to be lower by a factor of a few at $z>6$ \citep[e.g.,][]{Ouchi2010a, Jung2018a}; in Appendix \ref{sec:templates} we discuss how the inclusion of the full-strength Ly$\alpha$ templates marginally changes our solutions.
In order not to bias the redshift solutions of an intrinsically rare population at high redshift and in environments hypothesized to be overdense, we elect not to use a magnitude prior \citep[as has been done for other searches for EoR LBGs, e.g.,][]{Rojas-Ruiz2020a, Bagley2022a}.
We run EAZY for a range of redshifts from $0.05 < z < 12$ in steps of $\delta z = 0.05$. 

We use redshift probability density functions (PDFs) output by EAZY, where \pz\, $\propto \rm exp(-\chi^2/2)$.
LBG candidates are then classified as \textit{secure} or \textit{marginal} based on the EAZY fits. 
Photo--z fitting is not expected to have a precision much better than $\sigma_z \sim 0.5$, especially when dealing with a low number of broadband filters, so we begin with candidate LBGs with a best fit redshift $z_{phot}$ within $\Delta z = 0.7$ from the quasar's redshift.
We tested a number of $\Delta z$ criteria to verify this, finding that restricting to $\Delta z = 0.25$ yields 0--2 LBGs in 4/6 fields (or 2--3$\sigma$ below expectations from the blank field UVLF), indicating that our photo-$z$'s are not precise enough for a smaller redshift selection window.

Next, to eliminate candidates with a secondary low-redshift solution and ensure that \pz\, strongly prefers a high-redshift solution, we further impose the criterion that 70\% of the integral under \pz\, must be between $z_{phot} \pm 0.5$, which was the observed median $1\sigma$ margin of \pz\, in all SNR$>$3 sources in our sample.
In addition to this requirement, we require that 40\% of the integral under \pz\, must be within $z_{\rm qso} \pm 0.7$, i.e. there is at least a 40\% probability that the galaxy lies within the redshift bin broadly associated with the quasar rather than lying on the edge of the selection window.
Candidates are also required to have SNR $>$ 3 in both \hband\, and \jband\, (\jhband\, and \yjband) in the primary (parallel) fields, which helps to filter out spurious sources from the detection image (we justify this SNR cut in $\S$\ref{sec:completeness}). 

Finally, \vband\, and \iband\, fluxes must have low SNR to ensure a Lyman break at the correct redshift.
To ensure a robust dropout, we require \vband\, SNR $<3$, high enough to allow for some marginal emission that is possible if non-ionizing UV photons can still be transmitted through the Ly$\alpha$ forest \citep[e.g.,][]{Finkelstein2015a}; however, in practice, the median \vband\, SNR of sources that pass the above criteria is $0.5-1$ in all fields.
Last, we do not impose a strict criterion on the \iband\, SNR since the Lyman break falls in the middle of \iband\, between $6<z<7$; again, in practice the median \iband\, is $<2$ in all fields, but sources with marginal emission are evaluated individually as we outline in $\S$\ref{sec:lowz}.

To summarize, the following criteria are used to declare a candidate secure:

\begin{enumerate}
    \item 70\% of the integral under \pz\, is between $z_a \pm 0.5$.% ensuring that the PDF is not too broad.
    \item The best-fit redshift and 40\% of the integral under \pz\, is within $z_{\rm qso} \pm 0.7$.%, i.e. there is at least a 40\% probability the galaxy lies in the redshift bin of the UVLF.
    %\item Reduced $\chi^2 < 50$
    \item \hband\, and \jband\, (or \jhband\, and \yjband) fluxes have SNR $>$ 3.
    %\item \yband\, (\yjband) must have m$_{\rm AB} > 22$, which effectively eliminates main sequence stars. %is the limiting depth assumed for stars in the Gaia catalog. 
    \item \vband\, has SNR $\lessapprox$ 3.% to ensure a Lyman break.
\end{enumerate}

The marginal criteria are the same except for criterion 2, where the integral under \pz\, may be between 40\% and 70\% of the total. 
We impose an additional criterion to marginal sources: $\Sigma$P($z<2$)/$\Sigma$P($z>2$) $<$ 0.5.
This extra criterion allows us to include sources close to the detection limit which may have less constraining \pz, but filters out sources more consistent with red $z<2$ galaxy solutions that are degenerate with $z>6$ LBGs. 
\begin{figure*}
    \centering
    \includegraphics[width=2.0\columnwidth]{ 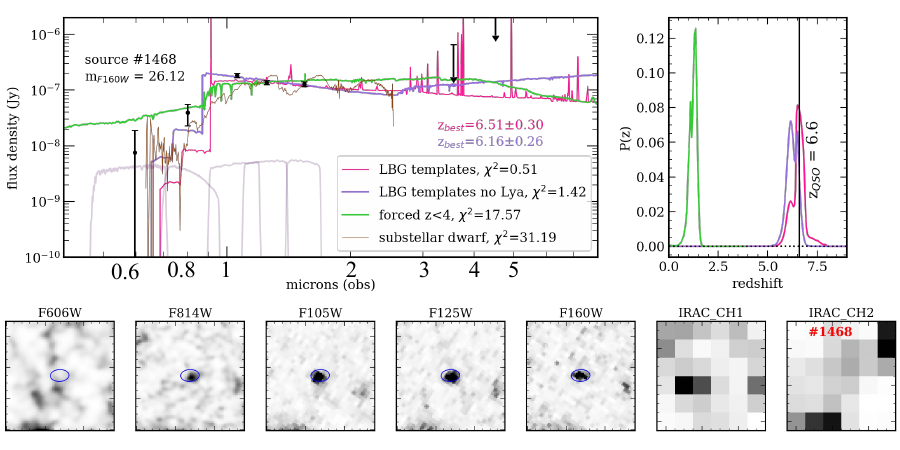}
    \caption{Top left: Example SED of a secure source in the J0305 field. Black points show photometry from \textit{HST} (with transmission curves on the bottom) and \textit{Spitzer}/IRAC 3.6\,\um\, and 4.5\,\um\, upper limits. The pink line shows the best fit EAZY SED containing a Ly$\alpha$ line with 100\% escape while the purple line forces a solution with reduced or fully extincted Ly$\alpha$: both fits are consistent with a $z\sim6$ solution within 1$\sigma$, but we eliminate the full-strength Ly$\alpha$ template for the redshift range we consider (see Appendix \ref{sec:templates}). The green line shows a fit forced to $z<4$, which is significantly worse than the $z>6$ solution as demonstrated by $\chi^2$. The brown line shows the best fit brown dwarf template spectrum from the SpeX Prism Library, showing that the solution is more consistent with a $z=6$ galaxy.  Top right: Normalized \pz\, distributions for the best fit LBG templates from EAZY, with the same color codes as the left panel. Bottom: 3.5\arcsec\, cutouts in all available filters; the blue ellipse indicates the Kron aperture in which photometry was extracted from the \textit{HST} images, set by the \hband\, detection image.}
    \label{fig:sed}
\end{figure*}

The initial selection returns 67 candidates in the J0305 quasar field (hereafter J0305 QSO), 42 candidates in the J0305 parallel (hereafter J0305 PAR), 72 in J2054 QSO, 57 in J2054 PAR, 26 in J2348 QSO, and 17 in J2348 PAR.
After these selection criteria were applied to all extracted sources, we then manually inspected the cutouts from \textit{HST} and \textit{Spitzer}, eliminating image artifacts, sources on the edge of the detector, and sources which fell in the ACS chip gap.
We manually eliminated a source if the implied rest-frame $M_{UV}$ was $<-23$ and sources were resolved beyond 0.3\arcsec since objects like these should be exceedingly rare at $6<z<7$, based on both the size-luminosity relationship \citep{Grazian2012a} and the UV luminosity function \citep{Finkelstein2016a}.
Objects were also eliminated if EAZY finds a redshift solution inconsistent with obvious emission in bands blueward of the expected dropout (i.e., $>3\sigma$).
We evaluate sources with marginal ($>2.5\sigma$) \iband\, detections on a case-by-case basis and discuss the process of vetting them in $\S$\ref{sec:contam}.
We use the best fit SED from EAZY to estimate the rest-frame absolute magnitude $M_{1450}$\, for each source by applying a tophat filter of width 100\AA\, centered at rest-frame 1450\,\AA. 

After applying the above criteria, we find a total of 46 sources (secure+marginal) in J0305 QSO, 20 in J0305 PAR, 23 in J2054 QSO, 10 in J2054 PAR, 6 in J2348 QSO, and 4 in J2348 PAR. 
Figure \ref{fig:sed} shows an example SED and cutout images of a secure candidate in the J0305 field.
The figure shows the difference between running EAZY with templates that force a Ly$\alpha$ emission line with 100\% escape versus only including the damped or fully extincted Ly$\alpha$ L22 templates, which result in slightly different best-fit redshifts but statistically consistent for the purposes of this study (see Appendix \ref{sec:templates}). 
We next discuss further constraints on contaminants to reduce our source list. 

\subsection{Selection Completeness and Contamination}\label{sec:completeness}
To estimate detection completeness and contamination, we construct 500 mock \hband\, maps with the same range of depths as the original images and inject sources similar to the expected properties of high-redshift LBGs. 
Specifically, we build noise maps for each filter with the same flux density distribution as the real maps, which is then convolved with the \hband\, PSF and rescaled to conserve the original RMS value in the PSF-matched maps.
We compute mock broadband photometry from the L22+EAZY templates in a redshift grid from to $1<z<8$, randomly perturbing the fluxes within our observed measurement uncertainty and scaling them to produce \hband\, fluxes between $23 < m < 28$.  
We randomly draw sources from this grid of mock photometry and assign the total fluxes to a 2D Gaussian with effective size $r_{\rm eff} = 0.2$\arcsec\,($\sim 1$\,kpc at $z=6.5$), based on the size-luminosity relation at $z=7$ \citep{Grazian2012a} .
These sources are convolved with the PSF and inserted at random positions in the mock noise map.
The sources are then extracted using the same SE parameters and photometry corrections as the real catalogs before being fed to EAZY. 
The extracted sources are considered a match if the output position is $<0.15$\arcsec\, from the input and $(z_{\rm phot}-z_{\rm input})/(1+z_{\rm input}) < 0.2$. 
We measure completeness as the ratio of matched output sources to input sources as a function of input absolute magnitude at $6<z_{\rm input}<7$.
Contamination is the number of extracted sources with  $(z_{phot}-z_{input})/(1+z_{input}) > 0.2$ in the same (input) redshift bin.
We find that we can reliably recover injected sources with \hband\, SNR = 3 since we assumed a blue spectral slope, i.e. they are detected at higher significance in \yband\, and \jband; thus we find that spurious sources (extracted ``sources" with no input match) never make it into our final photo-z selection at $M_{UV} < - 19$. 
We notice a systematic reduction in the total measured flux versus the input flux of 3--5\% even after the aperture correction, which we attribute to flux loss in the wings of the PSF, but we do not correct our real data for this since this translates to a negligible offset in magnitude.
Throughout the remainder of this paper, we will consider the UV luminosity function down to $M_{UV} = -19.5$, which is our 50\% completeness limit. 
We will later apply a fidelity correction as (1 -- contamination)/completeness as a function of absolute UV magnitude but not redshift, as we found it was basically constant between $6<z<7$.

\subsection{Sources of Contamination}\label{sec:contam}
\subsubsection{MLT Dwarfs}\label{sec:mlt}
The contamination rate from low-luminosity stars, i.e. MLT dwarfs, is expected to be low for fields at high galactic latitude.
Not coincidentally, J2054, the closest field to the Galactic plane at $b = -27^{\circ}$, had far more bright \textit{Gaia} stars than the other two fields and is therefore the most prone to substellar contaminants in LBG searches.
We calculated the expected number of MLT dwarfs in each field following the method of \citet{Euclid2019a}, where we assume that the Milky Way stellar density varies as a function of scale height $Z$, i.e. $\rho = \rho_0 e^{Z/Z_{\odot}}$.
We take $\rho_0$ to be the empirical density as a function of spectral type and absolute $J$ magnitude, taken from \citet{Bochanski2010a} and \citet{Best2021a}.
Transforming volume density to surface density, the expected number per arcmin$^2$ is expressed as:

\begin{equation}
    \frac{dN}{dJdA} =  \frac{ln 10}{5} \rho_0 e^{Z/Z_{\odot}} (10^{3(m-M+5)/5}) (\pi/180)^2
\end{equation}

where $\rho_0$ is in number pc$^{-3}$, $Z_{\odot}$ = 300 pc, $m$ is the observed apparent magnitude, and $M$ is absolute MKO $J$ magnitude (approximately the same as \textit{HST} F125W).
We then sum the expected contributions from each spectral type as a function of distance modulus in the range of apparent F125W magnitudes $22<m_J<27$, where we assume that a star with $m < 22$ would be in the \textit{Gaia} catalog (thus already eliminated) and $m > 27$ is below the 5$\sigma$ detection limit in the \textit{HST} data.
The expected total number of MLT stars is $\ll$ 1 in the J0305 and J2348 fields but $\sim$7 $\pm$ 3 in J2054, with the distribution peaking at $m_J$ = 25.5.
After examining the candidate sources which pass our LBG criteria, we flag any as stars based on compact circular morphology, \jband\, magnitude, and a simple minimized-$\chi^2$ comparison to empirical brown dwarf spectra from the SpeX Prism Library \citep{Burgasser2014a}.
We eliminate 3 sources between $23<m_J<25.5$ in the J2054 quasar field that passed our initial LBG color and photo-$z$ criteria, as they are more consistent with stellar templates ($\chi^2_{MLT} < 3$).
We also eliminate 3 sources from the J2054 parallel field, one candidate from the J0305 parallel field, and zero sources in the remainder of the fields, bringing the total counts to 46 in J0305 QSO, 19 in J0305 PAR, 20 in J2054 QSO, 7 in J2054 PAR, 6 in J2348 QSO, and 4 in J2348 PAR.

 \begin{figure*}
    \centering
    %trim goes {left bottom right top}
   \gridline{
   \includegraphics[trim={0.1in 0.05in 0.1in 0.05in}, width=0.69\columnwidth]{ 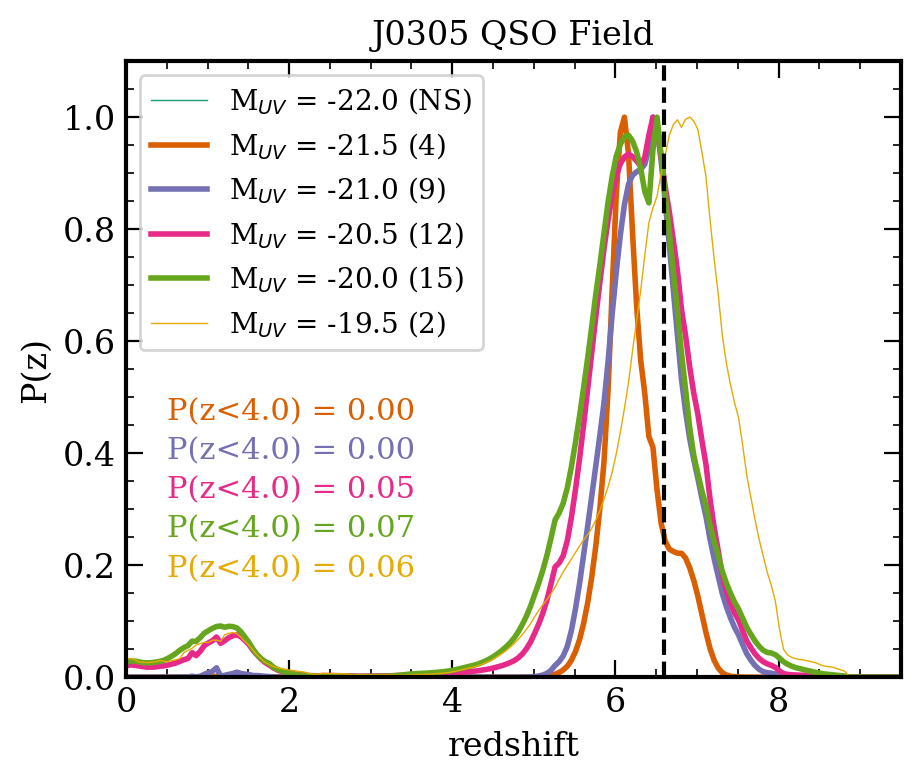}
   \includegraphics[trim={0.1in 0.05in 0.1in 0.05in},width=0.69\columnwidth]{ 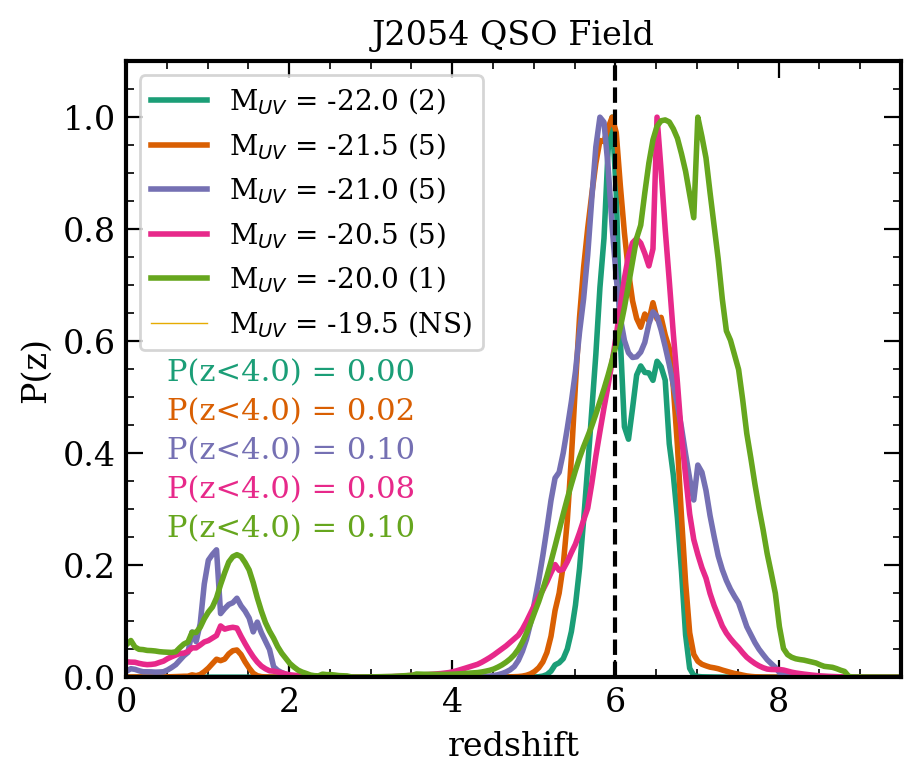}
   \includegraphics[trim={0.1in 0.05in 0.1in 0.05in},width=0.69\columnwidth]{ 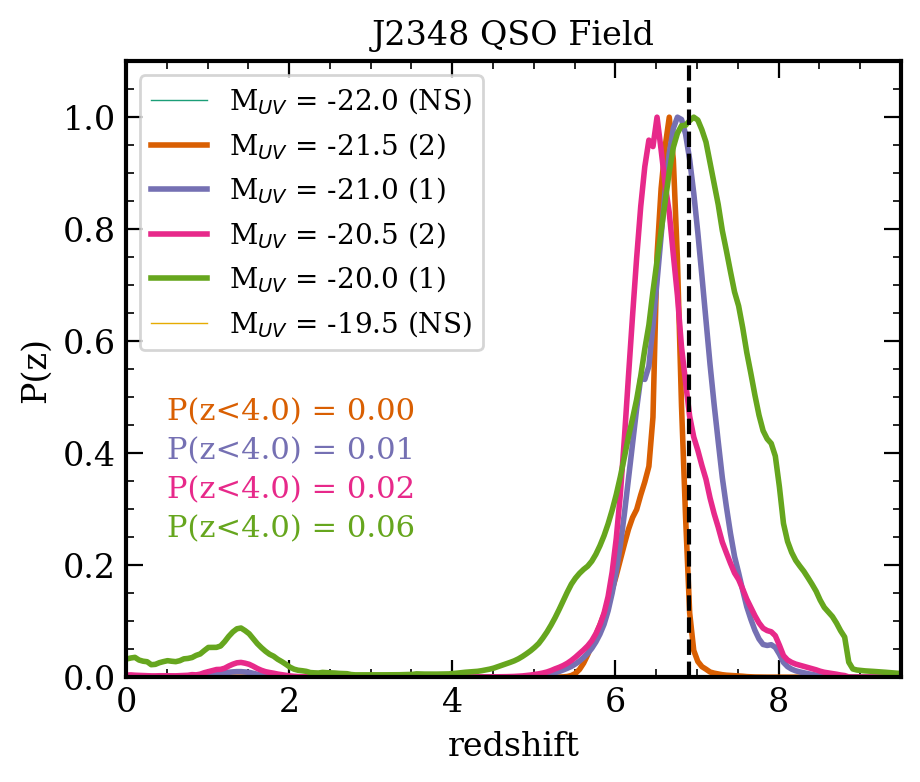}
   }
   \gridline{
   \includegraphics[trim={0.1in 0.05in 0.1in 0.05in},width=0.69\columnwidth]{ 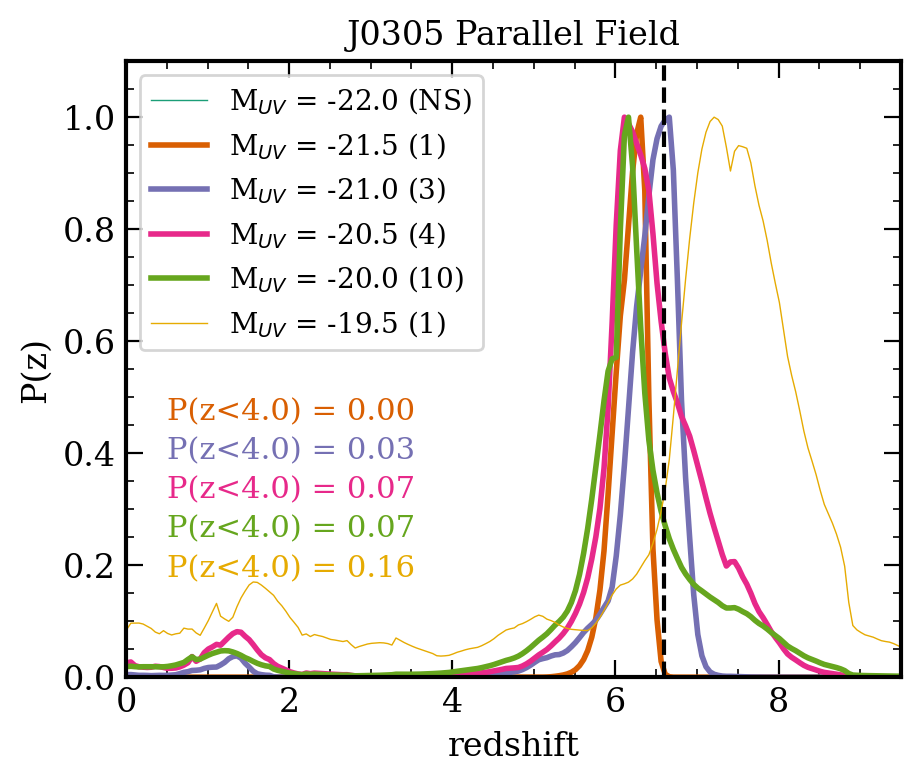}
   \includegraphics[trim={0.1in 0.05in 0.1in 0.05in},width=0.69\columnwidth]{  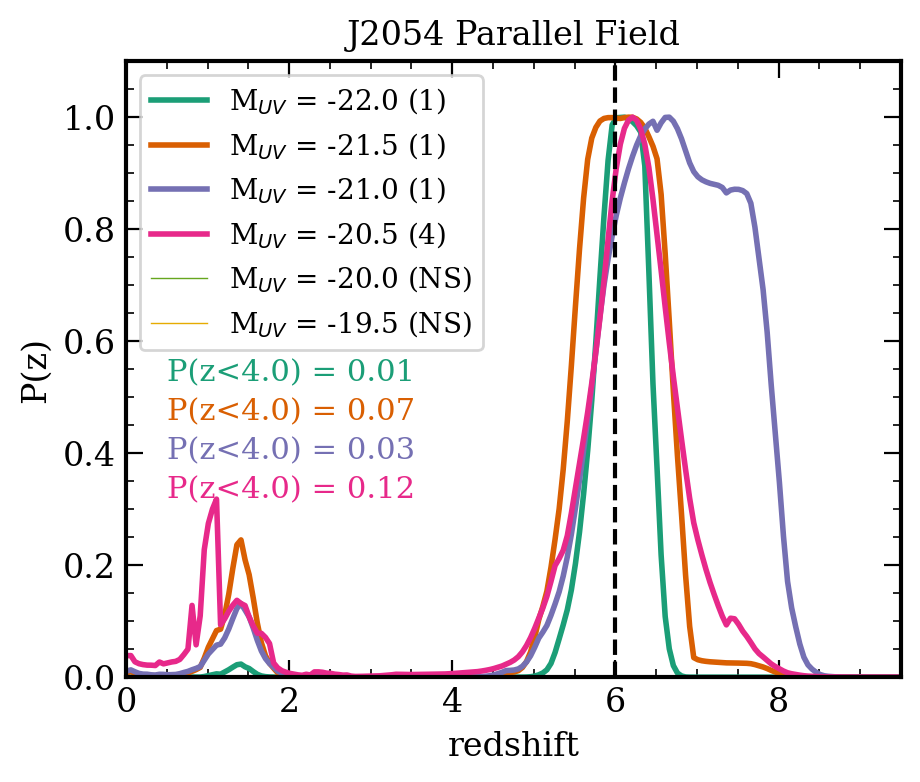}
   \includegraphics[trim={0.1in 0.05in 0.1in 0.05in},width=0.69\columnwidth]{ 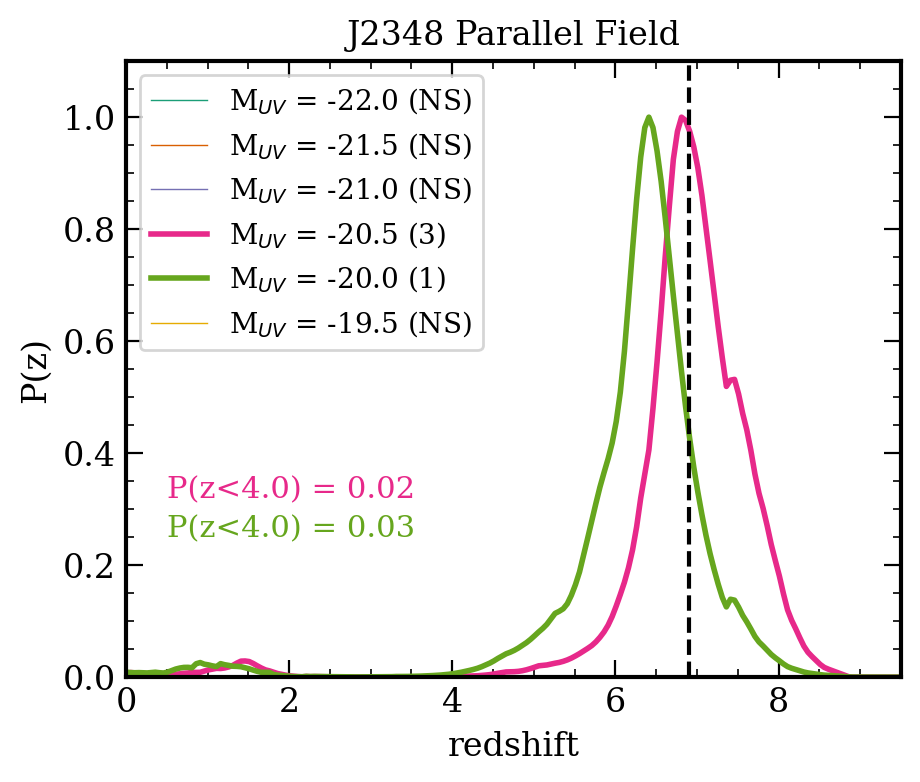}
   }
   
    \caption{Stacked EAZY \pz\, in bins of absolute magnitude M$_{UV}$ for each field, showing the level of contamination estimated purely from photometric redshift fitting. The legend denotes how many sources lie in that magnitude bin (with `NS' for no sources) and the black dashed line indicates the quasar redshift.  The inset text displays the fraction of the integrated total \pz\, with solutions of $z<4$, showing that our final selection criteria successfully includes only sources with a strongly preferred $z\sim6$ solution.}
    \label{fig:contam}
\end{figure*}

\subsubsection{Lower-Redshift Interlopers}\label{sec:lowz}
A full analysis of the contamination rate from lower-redshift galaxies would involve creating mock observations of a full suite of simulated galaxies with varying intrinsic properties (e.g., size, light profiles, etc) from $0<z<8$, which is beyond the scope of this work since we are evaluating a few single pointings.
Instead we describe briefly how we mitigate against obvious low-redshift contaminants not already discarded from visual inspection or the above procedures. 

The first issue concerns the discrepancy between color selection criteria and photo-$z$ estimates, owing mostly to marginal emission in the dropout bands.
All of the candidates are full optical dropouts in \vband\, except for one source in J0305 with $z_{phot} = 6.4$ and a $\sim3\sigma$ \vband\, detection, which may be indicative of non-ionizing UV photons not being fully absorbed by the IGM. 
In the next dropout band, we note that some sources have \iband\, emission at the $2.5-3.5\sigma$ level (especially those with fits at $z<6.5$ where the Lyman break lies in the middle of \iband), which leads to a dropout color slightly bluer than typical color cuts.
We note that 11/46 sources in the J0305 QSO field (1 secure, 10 marginal) formally fail the \citet{Bouwens2015a} \textit{iJH} color selection\footnote{This color selection was (\iband$-$\jband) $> X\, \wedge$ (\jband$-$\hband) $<$ 0.4 $\wedge$ (\iband$-$\jband) $>$\,2(\jband$-$\hband)$+X$ where $X=0.8$ at $z=6$ and $X=2.2$ at $z=7$.}, in addition to 5/19 (1 secure, 4 marginal) in J2054 and 1/6 (1 marginal) in J2348.
Of these 17 sources, 7 fully drop out in \iband\, while the other 10 show marginal emission; still, all of them pass our \pz\, criteria which filter out sources with low-redshift solutions, and none are consistent with brown dwarfs.
Because the Lyman break at this redshift could be degenerate with a Balmer break from evolved stellar populations at $z<2$, we ran EAZY again for these candidates and forced a fit at $z<4$.
We found that the difference between the original fit and the forced low-redshift fit, $\Delta \chi^2 = \chi^2_{z<4} - \chi^2_{orig}$, ranges from 3 to 34, i.e. $\chi^2$ is minimized for the higher-redshift solutions.
Recent work using \textit{JWST}/NIRCam to identify $z>9$ sources have employed the criterion of $\Delta\chi^2 > 4$  in order to prefer the higher-redshift solution \citep[e.g.][]{Fujimoto2022a, Finkelstein2022d}.
Following this, we discard 4 sources from J0305, 1 from J2054, and none from J2348, bringing the total source counts to 42, 18, and 6 in the primary fields. 
We calculate the equivalent color selection space in \textit{iYJJH} using the template SEDs of secure candidates in the quasar fields and apply this to the parallel fields, finding one source (in J0305 PAR) that fails. 
This source has $\Delta \chi^2 \sim 23$ so we keep it in our candidate sample.
In summary, we eliminate a total of 5 sources from the QSO fields which failed the \citet{Bouwens2015a} color selection criteria \textit{and} had $\chi^2_{z<4} - \chi^2_{orig} < 4$; none were eliminated from the parallel fields.

\begin{figure}
    \centering
    \includegraphics[width=1.0\columnwidth]{ 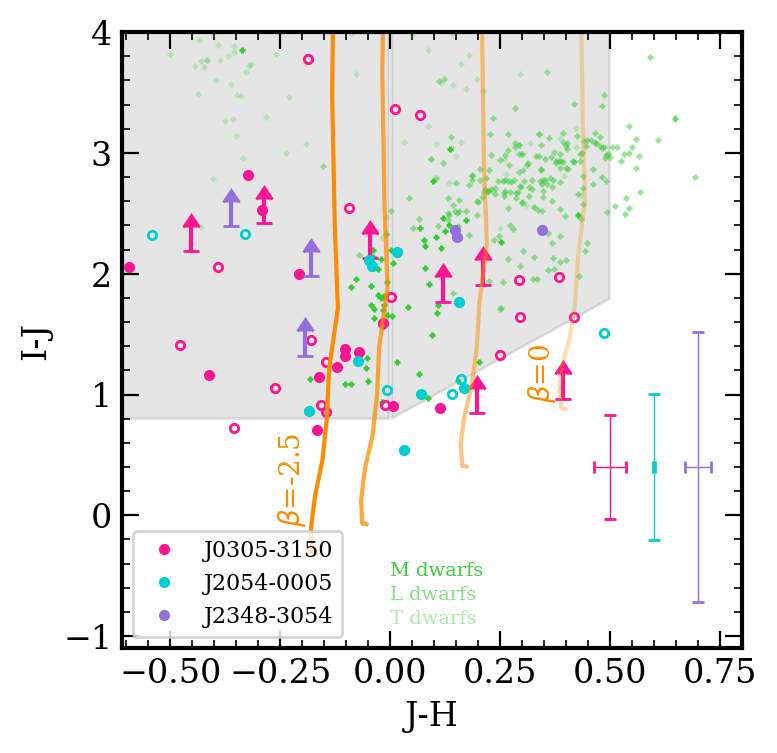}
    \caption{\iband--\jband\, vs. \jband--\hband\, for candidate LBGs in the three quasar fields. The solid colors show secure candidates in J0305--3150 (magenta), J2054--0005 (cyan), and J2348--3054 (purple) while empty circles indicate marginal candidates. Lower limit arrows on $I-J$ indicate where there is no detection in \iband. Gold, blue, and green crosses are colors of M, L, and T dwarfs respectively, calculated from the SpeX Prism Library \citep{Burgasser2014a}. The shaded region encompasses the $z=6$ selection criteria from \citet{Bouwens2015a}, where $J-H < 0.8$ and $I-J > 0.8$ and $I-J > 2(J-H) + 0.8$. The orange lines indicate the redshift tracks of a model LBG spectrum with UV slope $-2.5 < \beta < 0$, increasing left to right. Note that while several candidates appear to imply unusually blue continuum colors, all of the best fit EAZY templates are consistent with UV continuum slopes $\beta > -3$. The colored errorbars indicate the typical photometric errors in each field.}
    \label{fig:colors}
\end{figure}

Finally, we use the \pz\, distributions for our candidate sources to evaluate the level of remaining contamination from true lower-redshift sources.
In Figure \ref{fig:contam} we show the summed \pz\, distributions in each field as a function of absolute UV magnitude. 
We find that low-redshift contamination from brighter LBGs is low (0--10\%) while it rises up to 16\% for objects with $M_{UV} = -19.5$. 
We do not include sources with $M_{UV} > -19.5$ in our estimation of $\delta_{\rm gal}$ or the total \pz\, volume. 

In total, we end up with 42 sources in J0305 QSO (18 secure, 24 marginal); 19 sources in J0305 PAR (7, 12); 18 sources in J2054 QSO (9, 9); 7 sources in J2054 PAR (1, 6); 6 sources in J2348 QSO (4, 2); and 4 sources in J2348 PAR (1, 3). 
The positions and photometry of all sources are provided in Tables \ref{table:j0305q}$-$11.
Figure \ref{fig:colors} shows the \iband$-$\jband\, vs. \jband$-$\hband\, colors of the final candidates compared with observed brown dwarf colors from the SpeX Prism Library as well as the color selection criteria used by \citet{Bouwens2015a} for the blank field UV luminosity function in the COSMOS and CANDELS-UDS fields, showing that photo-$z$ fitting enables robust detections of galaxies that scatter just outside of the traditional galaxy color-color selection space.

%\section{Overdensity Results}\label{sec:discussion}
\section{Surface Density of LBGs}\label{sec:discussion2d}
\begin{figure*}
    \centering
    \includegraphics[width=0.95\textwidth]{ 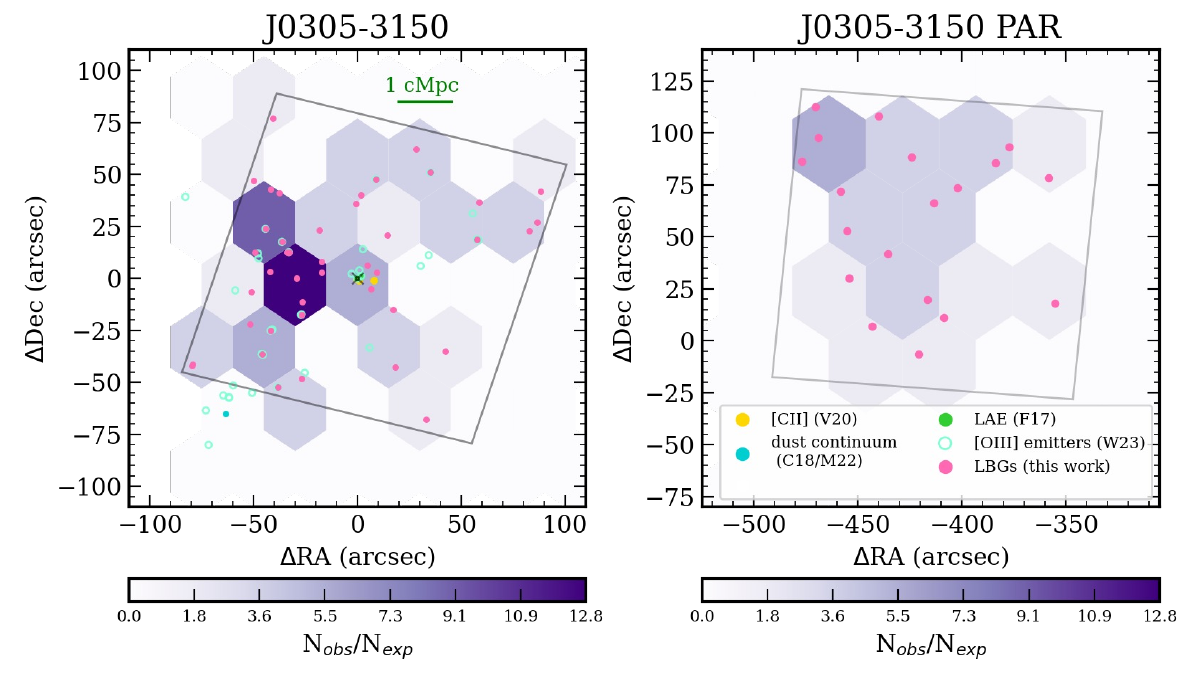}
    \caption{Surface density of LBGs in the J0305$-$3150 quasar field \textit{(left)} and parallel field \textit{(right)} compared with the number of expected sources based on the (completeness-adjusted) blank field UVLF within $\Delta z = 1$. We measure the density factor $N_{obs}/N_{exp}$ based on a grid with a bin size of 0.65 arcmin$^2$, shown by the colorbar. The gray solid rectangles show the WFC3 footprint. 
    The legend denotes other detections in this field: \Cii\, detections in yellow \citep{Venemans2020a}, continuum detections in teal \citep{Champagne2018a, Meyer2022a}, \Oiii\, emitters in light blue (Wang et al., submitted), an LAE in lime green and a Ly$\alpha$ blob coincident with the quasar host galaxy in darker green \citep{Farina2017a}. 
    The quasar is at $\Delta$RA, $\Delta$Dec = (0,0), shown by the black cross. 
    Note that the edge bins are not corrected for the fractional area outside of the field of view. 
    A strong 2D overdensity lies $\sim30$ arcsec ($\sim$1.2 comoving Mpc) away from the quasar.}
    \label{fig:j0305q-contours}
\end{figure*}

\begin{figure*}
    \centering
     \includegraphics[width=0.95\textwidth]{ 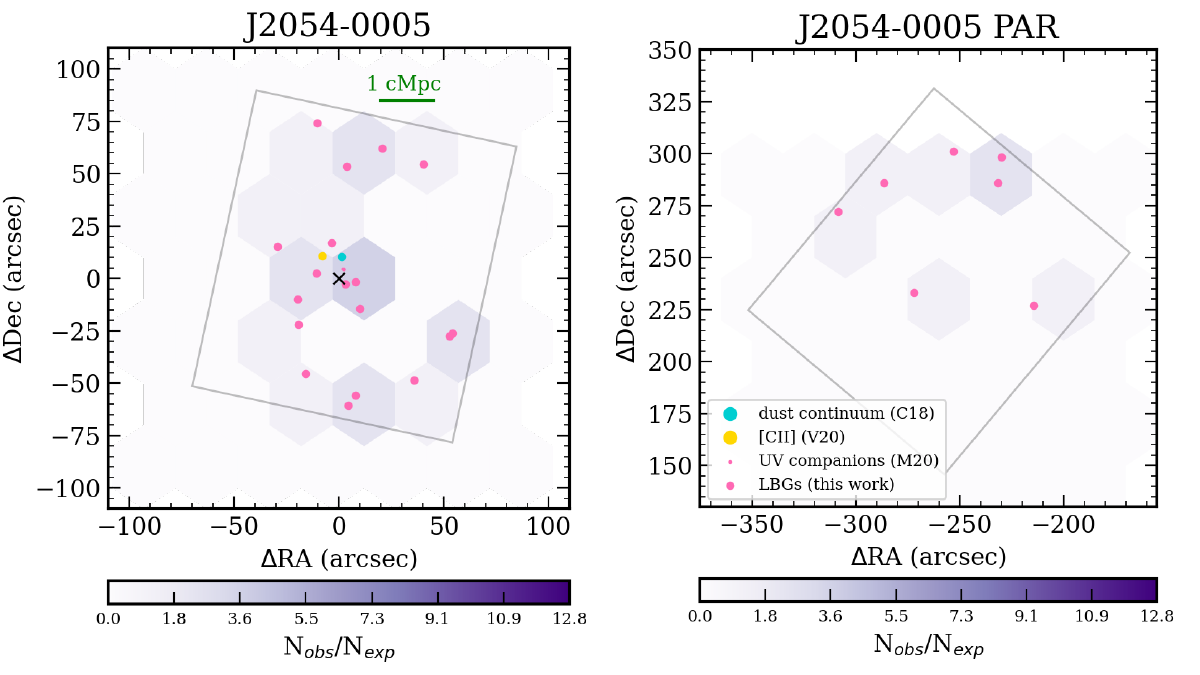}
     \caption{Same as Figure \ref{fig:j0305q-contours} for J2054$-$0005. 
     Small pink circles include UV galaxies identified with \hband\, and \jband\, data in \citet{Marshall2020a}.
     In two dimensions, there appears to be a modestly strong overdensity centrally located around the quasar, while the rest of the field is consistent with the blank field.}
    \label{fig:j2054q-contours}
\end{figure*}

\begin{figure*}
    \centering
    \includegraphics[width=1.0\textwidth]{ 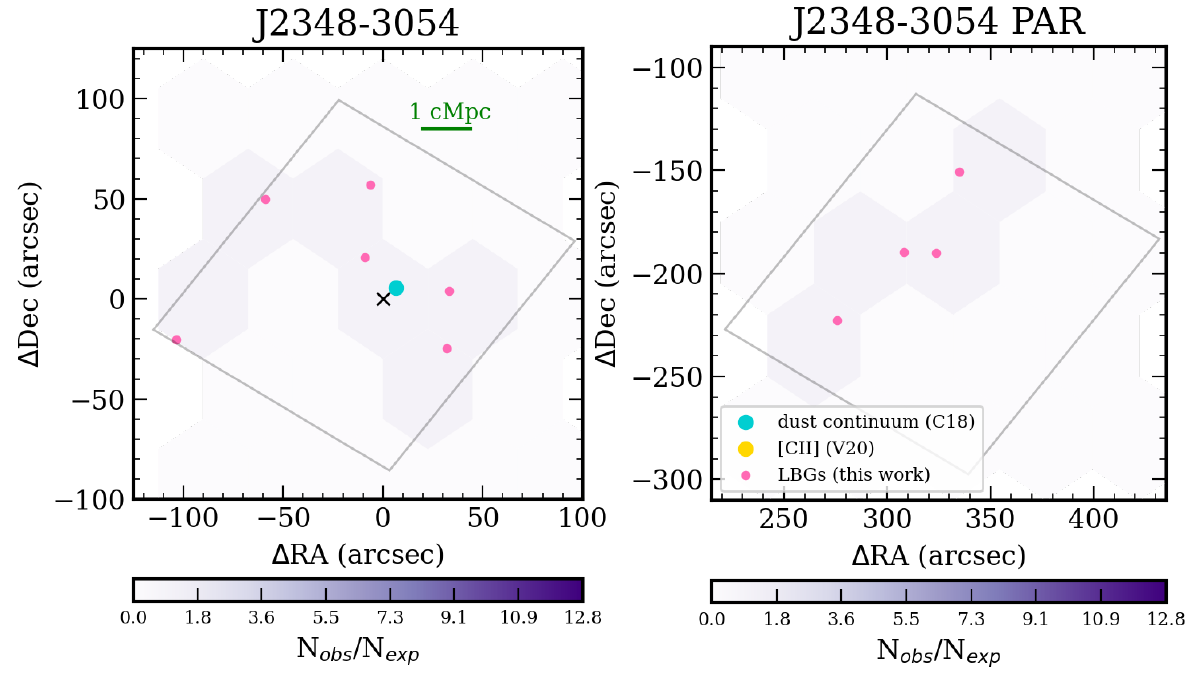}
     \caption{Same as Figure \ref{fig:j0305q-contours} for J2348$-$3054. Note that the color-scale has been normalized to the J0305 field for ease of visual comparison, but the area bins are larger, reflecting the declining number density of LBGs between $z=6$ and $z=7$. No overdensity exists in 2D. }
    \label{fig:j2348q-contours}
\end{figure*}

Before incorporating the 3-dimensional information offered by \pz, we first consider the two-dimensional distribution of LBG candidates compared to the average sky density.
In the absence of redshift constraints, we first consider the average sky density in each field by integrating the \citet{Finkelstein2016a} blank field UV luminosity function within a window of $\Delta z$ = 1 (see $\S$\ref{sec:uvlf} for more details), correcting for the observed completeness from our simulated maps.
We do not use the parallel fields as a control sample to measure the average surface density for two reasons: 1) the field of view is small, so the raw number counts are sensitive to cosmic variance, and 2) since they are separated by only $\sim 2$ proper Mpc from the primary fields, they may show signatures of the same overdensities if they exist in the quasar fields.
In Figures \ref{fig:j0305q-contours}, \ref{fig:j2054q-contours}, and \ref{fig:j2348q-contours} we show density maps of the LBGs in the three quasar fields and their parallels.
To quantify the local spatial distribution of the LBGs, we bin the detections on a hexagonal grid with a bin size of 30 arcseconds ($\sim$1.2 comoving Mpc) on each side, chosen to be large enough to encompass roughly one expected galaxy per solid angle bin.
We do not correct the bins on the edge of the field of view for the fractional area not observed, so we caution the reader not to interpret apparently truncated overdensities as real.

J0305 QSO shows an exceptional overdensity which could be distributed in a transverse filament to the east of the quasar, extending over 1 Mpc north-south.
J0305 QSO is already a well-studied field, containing LAEs, \Cii\, emitters, and \Oiii\, emitters, which we discuss in detail in the $\S$\ref{sec:j0305}.
Spectroscopic confirmation would be required to confirm whether the apparent filament is a chance alignment of LBGs along the line of sight, but overall it is clear that the overdensity is stronger to the east of the quasar. 
J0305 PAR also shows marginal evidence for local overdensities as well, though Poisson uncertainty dominates since most bins contain only 2 or 3 sources.

J2054 QSO suggests a modestly strong local overdensity (4$\times$ blank field expectations across multiple bins) of galaxies in the immediate vicinity of the quasar, all of which are secure detections.
However, the overdensity does not remain across the full field of view. 
\citet{Marshall2020a} examined this field using only the \jband\, and \hband\, data and reported 8 serendipitous detections within a projected distance of 20 kpc from the quasar, noting that 4 of them were consistent with being at lower redshift.
Of their 4 robust candidates at $M_{UV} < -20.5$, three of them are not in our catalog (likely due to different source detection strategies), and one of them had a best-fit redshift of $z_{phot} = 3.8$ in our sample.
Nonetheless, the three additional sources from that study strengthen the conclusion that a local overdensity is indeed physically associated with the quasar.
Its parallel field shows no deviation from the blank field expectation.

In the J2348 QSO field, we broaden the area bin to 45 arcseconds on a side because we expect a lower spatial density of LBGs at $z=7$. 
We see that in both the quasar and parallel fields there is no evidence for preferential enhancements of LBGs.

To quantify these apparent 2D overdensities in relation to the central quasar, we measure the overdensity, $\delta_{\rm gal} (r)$, as a function of separation from the quasar rather than in gridded area bins.
$\delta_{\rm gal}$ is defined as: 

\begin{equation}
    \delta_{\rm gal} = \frac{N_{obs}-N_{exp}}{N_{exp}}
\end{equation}

Here, $N_{exp}$ is the blank field expectation (corrected for the observed completeness), where here we adopt the convention of measuring sources within $\Delta z = 1$, and $N_{obs}$ is the raw number of LBGs in that same redshift bin. 
We account for incompleteness in the error, $\sigma_{\delta_{\rm gal}}$, by adding in quadrature the Poisson uncertainty and the fidelity results of our completeness simulations.
We also measure an error on $N_{exp}$ through a Monte Carlo sampling of the best-fit Schechter parameters to the parameterized $z=6-7$ UVLFs from \citet{Finkelstein2016a}. 

\begin{figure*}
    \centering
    \includegraphics[width=1.3\columnwidth]{ 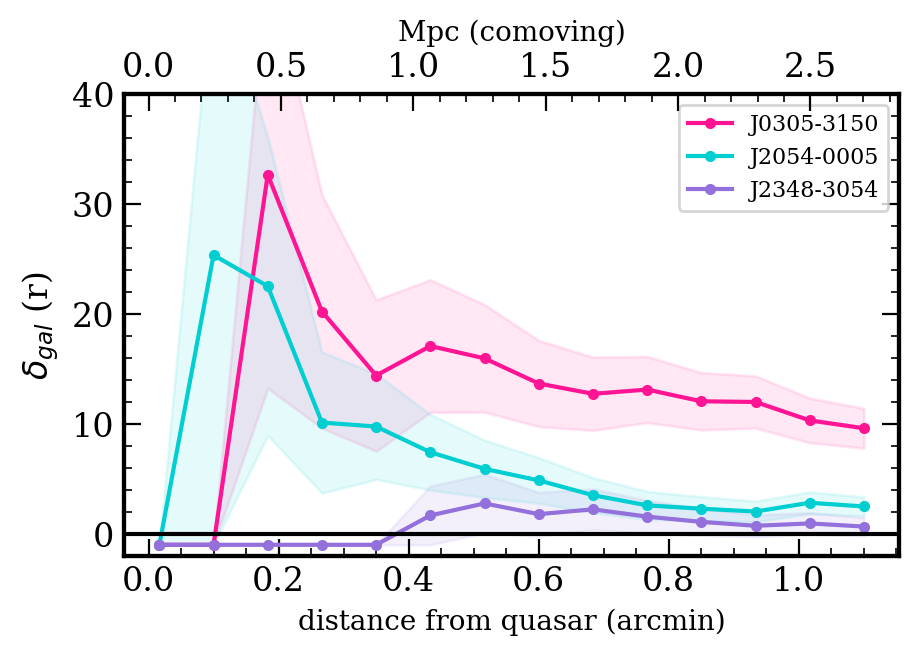}
    \caption{Overdensity $\delta_{\rm gal}$ as a function of angular separation from the quasar (transverse comoving units on top axis). Shaded regions indicate 1$\sigma$ uncertainty on $\delta_{\rm gal}$ which includes both Poisson uncertainty and cosmic variance. J0305 QSO shows a peaked overdensity 500 ckpc away but is enhanced at all separations, while J2054 QSO shows a strong overdensity at short separations that resembles the field at larger distances. J2348 QSO shows no evidence for a deviation from field expectations.}
    \label{fig:dgalr}
\end{figure*}

The result is plotted in Figure \ref{fig:dgalr}.
J0305 QSO showed a strong overdensity separated from the quasar, which we see reflected in the fact that the peak overdensity occurs 0.2\arcmin\, (0.5 comoving Mpc) away from the quasar.
If the 2D overdensity is real in 3D space, the quasar may not lie at its virial center.
As seen in Figure \ref{fig:j2054q-contours}, when the sky aperture is centered on the quasar, a strong peak with $\delta_{\rm gal} \sim 25$ arises very close to J2054, but it drops off quickly to $\delta_{\rm gal} = 1$ within 1 Mpc. 
This could indicate a compact overdensity physically related to the quasar, if confirmed to be real in 3D space. 
Finally, as expected from the 2D density maps, J2348 QSO remains consistent with the blank field at all separations. 
We note especially that measuring $\delta_{\rm gal}$ in apertures around the quasar yields a much stronger result than comparing to average projected 2D number counts, particularly in J2054.
While chance angular overdensities should be rare, all of the results centered on the quasar must be taken with caution since they are only evaluated as transverse separations: next we will use the combined \pz's to evaluate the strength of the overdensities in 3D space.

\section{3D Overdensities}\label{sec:3d}
\subsection{Consider the Volume}\label{sec:volume}
One of the most careful considerations we must make in declaring the existence of 3D overdensities is the effective volume probed by a search for clustered sources.
In the blank field, luminosity functions typically utilize the $V_{\rm max}$ method, which corresponds to the limiting volume in which a source could be observed in a given survey, magnitude bin, and redshift.
The effective volume $V_{\rm eff}$ is typically evaluated as the integral of the comoving volume element in a fixed redshift bin times a completeness selection function \citep[typically done via mock observations of LBGs as a function of luminosity, size and spectral slope, e.g.][]{Bagley2022a, Finkelstein2022a, Bouwens2022a}.
However, considering a fixed redshift range in a study targeting sources that are assumed to be clustered could change a true overdensity signal.
For instance, $\Delta z = 1$ within one WFC3 pointing spans a volume of ($\sim7000$\, Mpc)$^3$ at $z=6$, while the properties of protocluster cores at $z=6$ are typically evaluated in ($\sim$25 Mpc)$^3$ boxes \citep[e.g.,][]{Chiang2017a}.

In the ideal scenario of having spectroscopic confirmation of candidate companions, the volume spanned by a protocluster could be simply defined as the maximum redshift separation between members. 
But in the absence of spectroscopy, we can instead take advantage of the range of redshift solutions given by our photo-$z$ fitting method.
We compute the effective volume given the observed distribution of our sources, accounting for the uncertainty in \pz.
To do this, we co-add the individual \pz\, distributions of all LBG candidates, weighted by the fraction of their integrated \pz\, that lies between the prescribed $z \pm 0.7$ from the quasar redshift (see $\S$\ref{sec:eazy}) to obtain $\Sigma P(z)$.
This has the effect of suppressing sources with a lower probability of actually lying near the quasar and/or higher uncertainty in their redshift solutions, and allows us to evaluate the most probable volume spanned by clustered sources.
Because we restricted secure sources to have 70\% of the integrated \pz\, within $z_{\rm qso}\pm0.7$, the effective volume should be substantially narrower than the fixed $\Delta z$ = 1.4 if the sources are clustered.

\begin{figure*}
    \centering
   \includegraphics[width=2.0\columnwidth]{ 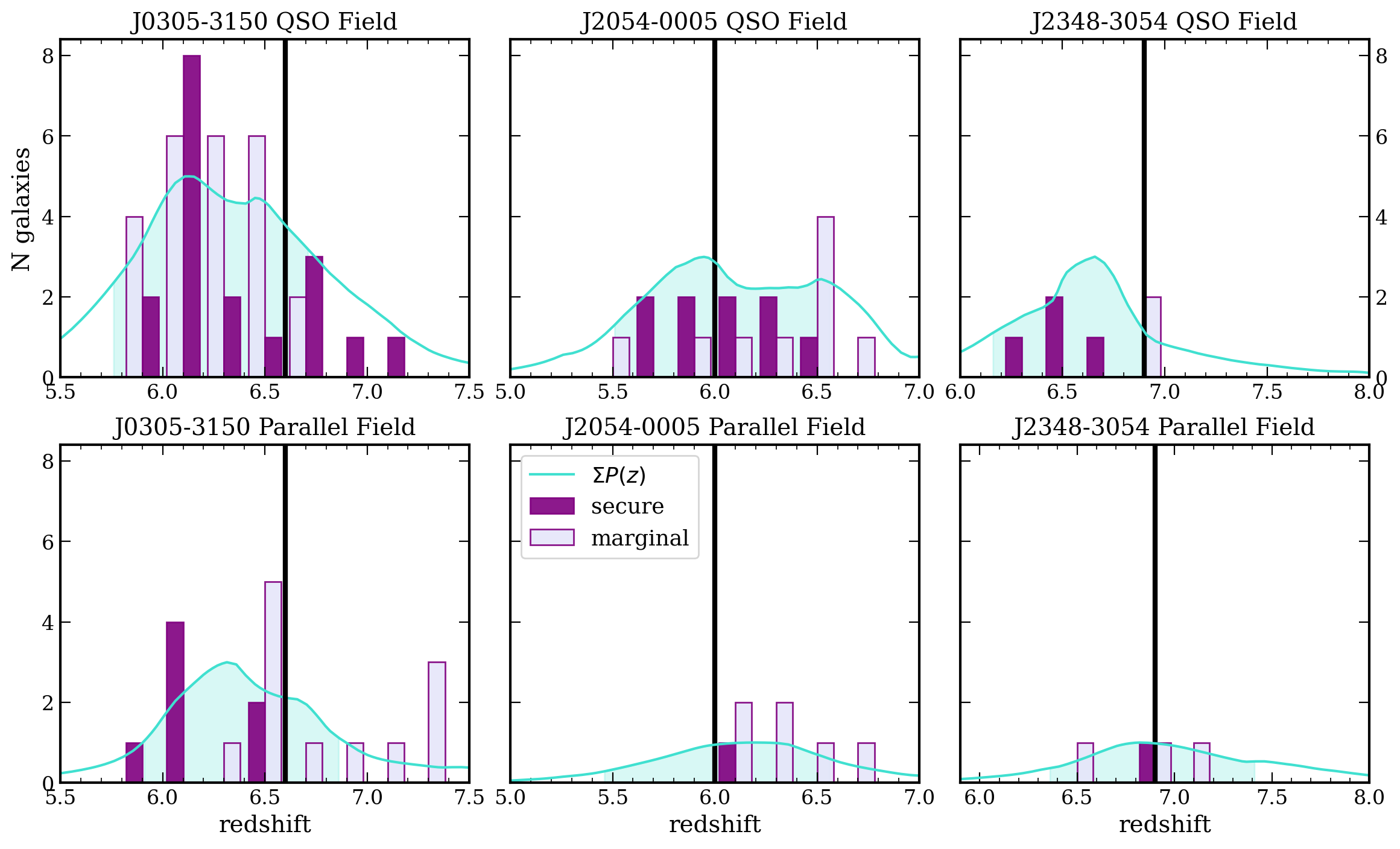}
    \caption{Redshift histograms of the LBG candidates, with secure candidates in dark purple and marginal candidates in light purple. The cyan lines represent the total $\Sigma P(z)$, the sum of individual cumulative \pz\, distributions of all (secure and marginal) LBG candidates, weighted by the percentage of the integrated \pz\, that falls within $\Delta z = 0.7$ from the quasar redshift (thick black line) --- effectively upweighting sources with a higher probability of lying near the quasar. The shaded area represents the 16th and 84th percentiles of the $\Sigma P(z)$ distribution from which the effective volume $\Delta V$ is calculated.}
    \label{fig:volume}
\end{figure*}

In Figure \ref{fig:volume} we show the histograms of best-fit redshifts for the secure and marginal sources, and demonstrate this method of co-adding the volumes with the limiting redshifts as the 16th and 84th percentiles of the cumulative $\Sigma P(z)$.
We can see how the effective volume incorporates the uncertainty in photometric redshift --- on one hand,  we wash out a real overdensity signal if the best-fit redshifts are accurate, but on the other hand it prevents us from overstating the significance of the overdensity if we were to use the minimum and maximum photo-$z$.
Using our method, the median $\Delta z$ for all fields is 1.1, ranging from 0.8--1.2.
To investigate whether our somewhat arbitrary selection choice of $z_{qso} \pm 0.7$ has an effect on our final result, we tested this calculation on source lists within $z_{qso} \pm 0.25, 0.5$ and $1$ and find that the effective $dz$ does not change from $\approx$ 1 since 1) this represents the typical spread of \pz, and 2) sources near the edge of the $z$ selection window are already down-weighted by our method of coaddition.\footnote{
Technically, $\Sigma P(z)$ does change when evaluated in separate magnitude bins (see Figure \ref{fig:colors}), but this is mostly a selection effect since the photo-$z$ fit is sensitive to the SNR of the photometry and is therefore inherently more uncertain for fainter galaxies.
Thus, in order not to bias the shape of the final UVLF, we use a constant effective volume in each bin.}
For the remainder of this study, we estimate the expected number of LBGs from blank field UVLFs within the volume of the WFC3 field of view (162\arcsec$\times$162\arcsec) $\times$ the effective $dz$ in each field. 

\subsection{Comparing to the UV Luminosity Function}\label{sec:uvlf}
We next calculate the UV luminosity function for our candidate LBGs.
While the blank field UV luminosity function has shown some evidence favoring a double power law rather than a Schechter function in recent ultra-high redshift ($z>8$) observations \citep[e.g.,][]{Bowler2020a, Bouwens2022c, Harikane2022a, Bagley2022a, Finkelstein2022c}, we assume the log Schechter fit is a valid comparison at $z=6-7$, expressed as:

\begin{equation}
    \phi(M) = 0.4\,\rm ln(10)\, \phi^*\, 10^{-0.4(M-M^*)(\alpha+1)}\, e^{-10^{-0.4(M-M^*)}}
\end{equation}

where $\phi(M)$ is in units of N mag$^{-1}$ Mpc$^{-3}$.  
We use the reference luminosity function from \citet{Finkelstein2016a} which expresses the Schechter parameters and their uncertainties as a function of redshift, i.e., $M^*(z)$, $\alpha(z)$, and $\rm log\phi^*(z)$ (see their $\S$5.3). 
We bin our galaxies with $\Delta$\,mag = 0.5 from $M_{UV} = -22.5$ to $-19.5$.
Each bin contains $N$ galaxies times the fraction of their integrated \pz\, that falls within the $1\sigma$ from the median of $\Sigma P(z)$ such that they are weighted by their probability of lying near the quasar, then divided by the volume we measured in $\S$\ref{sec:volume}.
We also corrected for the values of completeness and contamination in each magnitude bin from our simulations (see $\S$\ref{sec:completeness}), which we take to be constant in redshift space.
The uncertainty is measured as Poisson noise added in quadrature with the uncertainties of the completeness and contamination simulations.

\begin{figure*}
    \centering
   \gridline{ {\includegraphics[width=1.75\columnwidth]{ 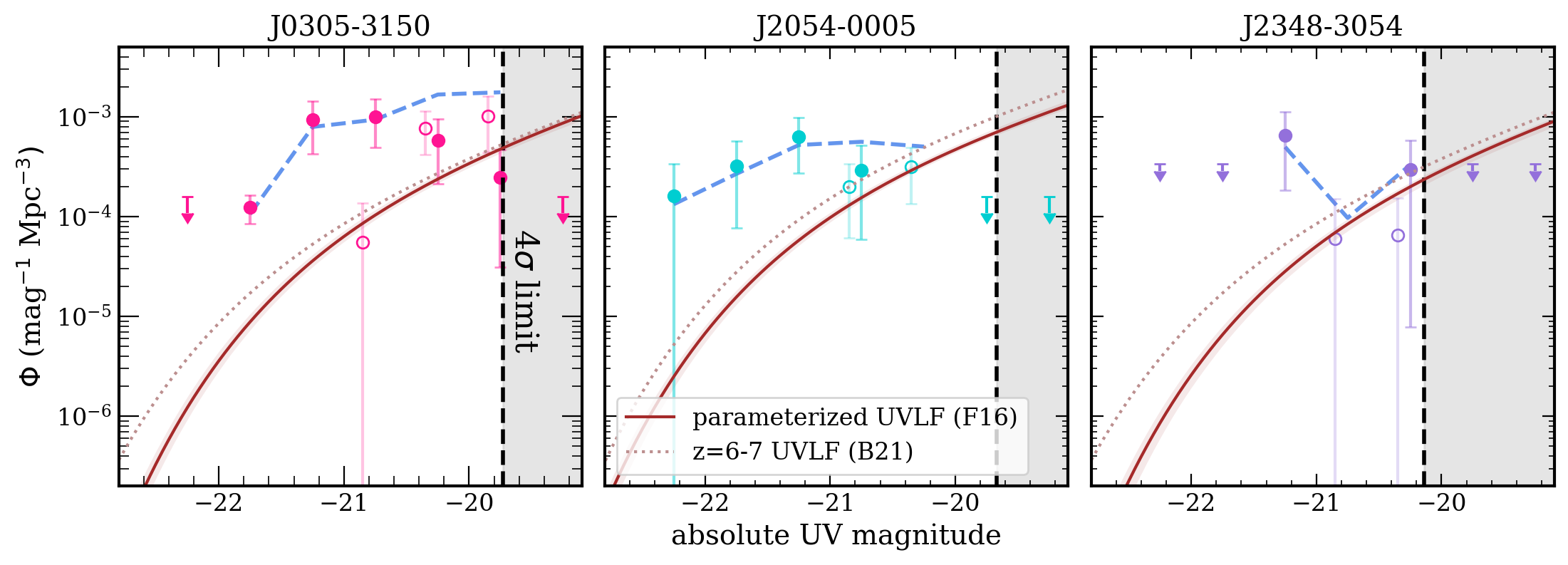}}
   }
   \vspace{-0.2cm}
   \gridline{
   {\includegraphics[width=1.75\columnwidth]{ 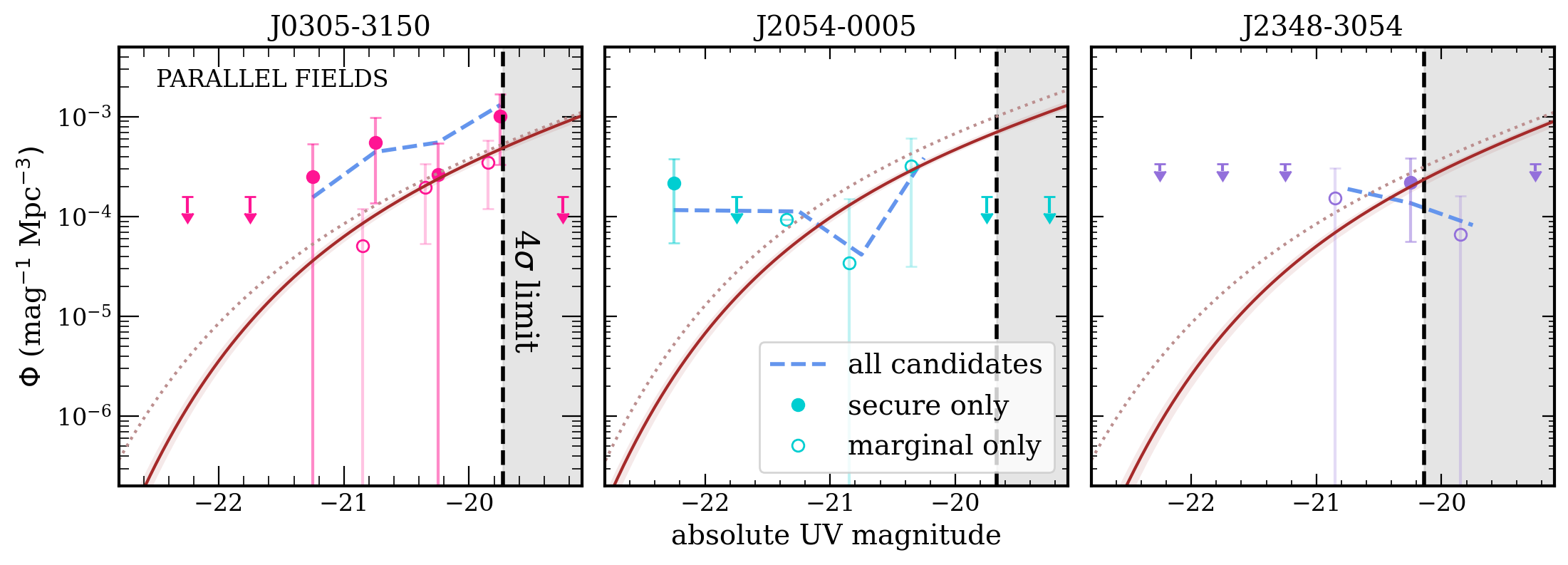}}
   }
    \caption{Individual UVLFs calculated for each quasar field (top) and parallel field (bottom), corrected for completeness and contamination. 
    The solid brown lines indicate the parameterized UVLF from \citet{Finkelstein2016a} evaluated at the quasar redshift, and the dotted brown lines indicate the $z=6$ (for J2054) or $z=7$ (for J0305 and J2348) UVLF from \citet{Bouwens2021a}.
    The filled circles show only \textit{secure} sources within the volume spanned by the 16th-84th percentiles of $\Sigma P_{sec}(z)$.
    The empty circles show only \textit{marginal} sources within the volume spanned by the same region of $\Sigma P_{marg}(z)$, offset slightly for visual clarity.
    The dashed blue line shows the sum of both populations, showing that the \textit{marginal} sources are necessary to populate the faint end of the UVLF even after correcting for incompleteness. 
    Our completeness simulations successfully recover sources with $M_{UV} < -19.5$ classified as ``marginal" with low ($<10$\%) contamination; therefore, we consider all LBG samples as secure+marginal.
    \label{fig:uvlfindivid}
    }
\end{figure*}

Figure \ref{fig:uvlfindivid} shows the luminosity functions for each field individually, where we have calculated $\Sigma P(z)$ and $\phi (M)$ separately for the secure and marginal source lists. 
We can see that the inclusion of the ``marginal" sources is necessary in order to reproduce the expected faint end of the blank field UVLF, since sources fainter than $M_{UV} = -20$ tend not to be classified as secure by our criteria.
On the other hand, because we have defined the marginal selection criteria to allow for relatively wide \pz's, their corresponding effective volume is much wider than for the secure sources, diluting the volume density at the faint end.
Since we have taken care to exclude low-redshift contaminants even from the ``marginal" sample, in the remainder of our calculations, we treat the secure + marginal source lists as one, but we note that $\Sigma P(z)$ is dominated by wider individual \pz's from the marginal list.

\begin{figure*}
    \centering
   \includegraphics[width=2.1\columnwidth]{ 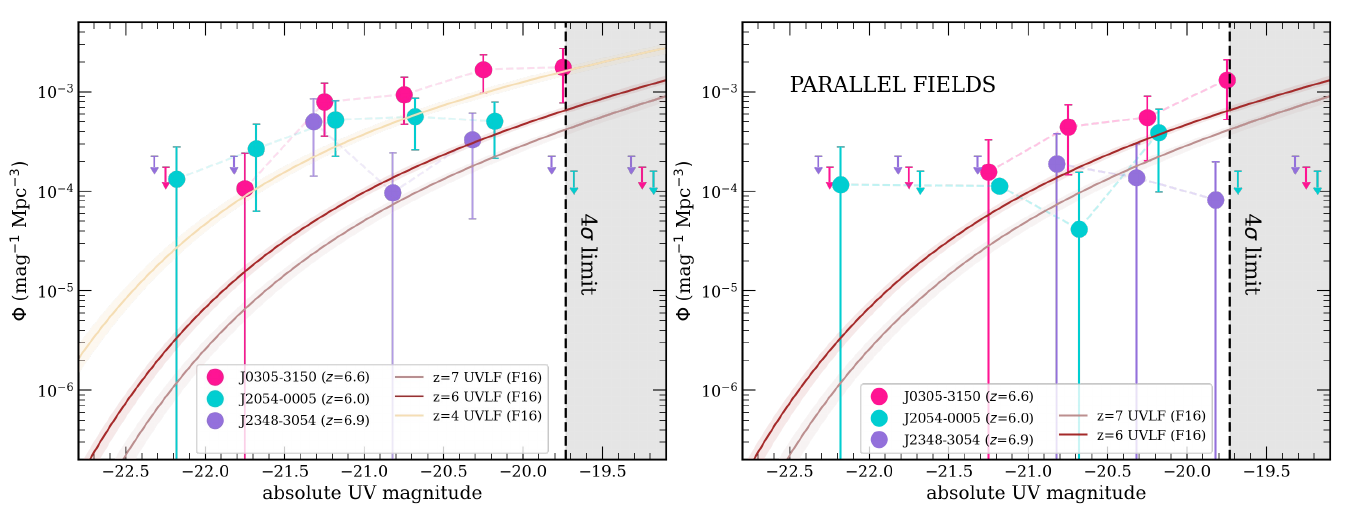}
    \caption{Luminosity function of the LBG candidates in the QSO fields (left) and parallel fields (right). 
     The three fields are shown in pink (J0305), cyan (J2054) and purple (J2348) in both panels, with points slightly offset on the $x$-axis for clarity. 
     The solid brown lines show the $z=6$ and $z=7$ blank field UVLF from \citet{Finkelstein2016a}, with shaded regions indicating 1$\sigma$ errors from an MC sample of the best fit parameters.
    The pale yellow line shows the $z=4$ UVLF from \citet{Finkelstein2016a}, the shape of which is most consistent with the most overdense field, J0305-3150.
    The black shaded region indicates the 4$\sigma$ limit where we reach 50\% completeness; thus the overdensity signal is considered only at magnitudes brighter than this bin.
    Note that the galaxies in each magnitude bin have been fractionally weighted by their \pz, such that this is the most conservative estimate of the luminosity function in these fields.}
    \label{fig:uvlf}
\end{figure*}

Figure \ref{fig:uvlf} shows the luminosity functions of the candidate LBGs in this sample, which have been weighted by the aggregate \pz, as well as comparisons to other literature UVLFs.
An overdensity is visible in all luminosity bins in the J0305 field, while J2054 is marginally consistent with an overdensity at $M<M_*$ and J2348 is consistent with the $z=7$ field UVLF in all luminosity bins.
J0305 and J2054 are statistically consistent with enhanced UVLFs associated with the quasar, since each LBG sample has been considered as fractional contributions to the UVLF within $z_{qso} \pm 0.7$.

Interestingly, when taking into account the uncertainty in \pz\, as well as the full field of view, the apparent local overdensity around J2054 is washed out except for marginal evidence for an overdensity at the bright end ($M<-21$).
While the overall shape of the UVLF in J0305 is consistent with that of the blank field, we note that it also appears to be stronger for sources at $M<M_*$. 
This could be explained partially by incompleteness, but we have corrected for this effect using completeness simulations, and the parallel fields show reasonable agreement with faint-end expectations given our depth.
We hypothesize that, if it is an observational effect, $\Sigma P(z)$ has been over-estimated for the fainter galaxies with more uncertain $z_{phot}$.
On the other hand, in particular for J0305 where there exists an overdensity even in the faintest bin, the comparatively lower overdensity at the fainter end could be physically motivated, as we will discuss further later. 
In any case, the overall normalization of the J0305 UVLF implies an overdensity 10$\times$ higher than the blank field UVLF --- this is the \textit{most} conservative estimate because we have weighted the observed UVLF by the photometric redshift uncertainties and thus could miss real faint galaxies.

We compare our observed number counts (not weighted by \pz\, as above) in 3D with \citet{Finkelstein2016a} and quantify the overdensity over the absolute magnitude range $-22<M<-19.5$ (see $\S$\ref{sec:discussion2d} for a similar calculation in 2D only).
This time we measure $N_{exp}$ by integrating the UV luminosity function across the WFC3 field of view and within the 1$\sigma$ range of $\Sigma P(z)$. 
We assume Poisson errors for $N_{obs}$ but we take into account both the Poisson uncertainty and cosmic variance in the error on $N_{exp}$ using the cosmic variance calculator\footnote{\url{https://www.ph.unimelb.edu.au/~mtrenti/cvc/CosmicVariance.html}} \citep{Trenti2008a}, which we estimate to be about 30\% at $z=6$ and 35\% at $z=7$.
We measure $\delta_{\rm gal}$ down to $M<-19.5$, which is our 50\% completeness limit.

In total, we detect 42 LBGs in the J0305 field versus $4.3 \pm 0.6$ expected from the literature UVLF evaluated at $z=6.6$ and corrected for the total completeness measured from our simulations (85\%).
J0305 results in the strongest overdensity, with $\delta_{\rm gal} = 8.8 \pm 1.8$.
J2054 shows a marginal ($2.7\sigma$) overdensity of $\delta_{\rm gal} = 1.9 \pm 0.7$, based on 18 total LBGs versus $6.3 \pm 0.5$ expected in our observations at $z=6$.
J2348 is consistent with the blank field, with $\delta_{\rm gal} = 1.3 \pm 1.2$, based on 6 candidates versus $2.6 \pm 0.7$ expected at $z=6.9$.
In a single \textit{HST} pointing with high completeness at $M_{UV} = -20$, the uncertainty due to cosmic variance is only slightly lower than the Poisson uncertainty, making it difficult to declare the overdensity $\delta_{\rm gal}$ to be at very high significance; however, J0305 still stands out as a very strong overdensity.

J0305 PAR, as seen in the 2D distribution, is consistent with an overdensity at the 3$\sigma$ level, with $\delta_{\rm gal} = 3.6 \pm 1.2$ based on 19 observed sources.
J2054 PAR and J2348 PAR are fully consistent with expectations at $\delta_{\rm gal} = 0.1 \pm 0.4$ based on 7 sources and $\delta_{\rm gal} = 0.0 \pm 0.6$ based on 4 sources, respectively.
Note that the LBGs in the parallel fields are selected in the same redshift window relative to $z_{qso}$ as the quasar fields, but $\Sigma P(z)$ is uniquely determined by the observed LBG distribution in each field; however, if we apply $\Sigma P(z)$ from the quasar fields to their respective parallels, the $\delta_{\rm gal}$ results are statistically identical.

\begin{deluxetable}{ccccccc}
\tablecaption{\textbf{Overdensity estimates.} We include the number of secure and marginal galaxies and their corresponding estimates of $\delta_{\rm gal}$. %for the {\tt ALL\_TEMPS} and {\tt NO\_LYA} runs from EAZY (see text for details).
$\langle z \rangle$ refers to the median best fit redshift and the 1$\sigma$ width of the co-added $\Sigma P(z)$. \label{table:dgal}}
\tablehead{\colhead{Field} & \colhead{N$_{sec}$} & \colhead{N$_{marg}$} & \colhead{N$_{tot}$} & \colhead{$\langle z \rangle$} & \colhead{$\delta_{\rm gal}$} 
}
\startdata
J0305 QSO & 18 & 24 & 42 & 6.3 $\pm$ 0.5 & 8.8 $\pm$ 1.8 \\
J2054 QSO & 9 & 9 & 18 & 6.2 $\pm$ 0.6 & 1.9 $\pm$ 0.7 \\
J2348 QSO & 4 & 2 & 6 & 6.6 $\pm$ 0.4 & 1.3 $\pm$ 1.2 \\
J0305 PAR & 7 & 12 & 19 & 6.5 $\pm$ 0.5 & 3.6 $\pm$ 1.2 \\
J2054 PAR & 1 & 6 & 7 & 6.2 $\pm$ 0.6 & 0.1 $\pm$ 0.4 \\
J2348 PAR & 1 & 3 & 4 & 6.9 $\pm$ 0.6 & 0.0 $\pm$ 0.6 
\enddata
\end{deluxetable}

\section{Discussion}\label{sec:discussion}

In this work, we have found three different distributions of galaxies around quasars towards the end of the EoR.
J0305 hosts a statistically significant overdensity across the full field of view, and there may be filamentary structure separated by a few hundred kpc.
J2054 shows a significant 2D overdensity ($\sim25\times$ the blank field at small angular separations from the quasar), but is overall still marginally consistent with the field in 3 dimensions.
J2348 shows no evidence for a 2D or 3D enhancement of LBGs.
Here, we explore some of the ways we can explore various physical scenarios that might give rise to the environments we have seen.

First, does the projected spatial distribution of LBGs imply a direct physical association with the quasar?
In the following, we will examine the dependence of LBG luminosity on the transverse separation from the quasar in overdense fields. 
In a blank field, one would expect a random spatial distribution of LBGs as a function of luminosity.
Despite strong field-to-field variation and the comparatively lower contrast above the background distribution even for truly clustered populations at $z>6$, one strong indication of a physically associated overdensity might be preferential enhancements of bright galaxies at close angular separations from the quasar.
For instance, ionizing radiation originating from quasars within the so-called proximity zone affects galaxies in its vicinity, which may preferentially suppress star formation in less luminous galaxies on $\sim3-5$ pMpc scales \citep[e.g.,][]{Eilers2017a, Bosman2020a}.

Another question we seek to answer below is whether or not we are dealing with observational biases (i.e., small field of view and large redshift uncertainty) that are diluting an ubiquitous overdensity signal or if the heterogeneity of environments is expected from models. 
The active phase of a quasar relatively short at $<10^7$ years \citep{Satyavolu2022a}, so this stochasticity compared to the timescale of enhanced star formation could mean we do not always observe an overdensity during the quasar's lifetime.
Finally, from a theoretical standpoint, we explore how to quantify the significance of an observational overdensity when comparing to expectations from dark matter simulations.

\subsection{The Nature of Individual Quasar Environments}\label{sec:nature}
We first discuss the implications of the spatial and luminosity distributions of the LBGs in each quasar field before placing them in multiwavelength context in $\S$\ref{sec:dsfg}.

\subsubsection{Evaluating the significance of the J0305 overdensity}\label{sec:j0305} 
The host galaxy and surrounding environment of J0305--3150 has been studied extensively albeit with varying spatial scales, depths, and tracers, including with ALMA, Subaru Suprime-Cam, VLT MUSE, and JCMT SCUBA-2.
At long wavelengths, one spectroscopically confirmed dust continuum source was found 7\arcsec\, away from the quasar \citep{Champagne2018a}.
ALMA followup of SCUBA-2--selected SMGs showed 2 more sources $\sim2$\arcmin\, away \citep{Li2020a}, representing no overdensity of continuum sources in the probed area.
In contrast, three \Cii\, sources were found $0.5-7$\arcsec\, away, two orders of magnitude above the volume density expected from blank field \Cii\, counts \citep{Gonzalez-Lopez2020a}.
\citet{Farina2017a} found an extended ($\sim$9\,kpc) Ly$\alpha$ blob is coincident with the host galaxy with a LAE 2\arcsec\, away, suggesting nearby mergers in an overdense environment.
Subaru Suprime-Cam narrowband searches for LAEs and LBGs across 700 arcmin$^2$ have arrived at a $1\sigma$ underdensity and $4\sigma$ overdensity respectively \citep{Ota2018a}.

Finally, Wang et al. (submitted) show a spectroscopic $\delta_{\rm gal} \sim 13$ overdensity of H$\beta$+\Oiii\, emitters identified with \textit{JWST}/NIRCam WFSS; they find 41 galaxies at $z>5.4$, 13 of which are within $\Delta z = 0.03$ from the quasar.
Of these 41, 15 fall out of the WFC3 footprint and 7 are too faint to have visible counterparts in the \textit{HST} data, since the \textit{JWST} data finds the faintest \Oiii\, emitters at $2 \times 10^{-18}$\, erg\,s$^{-1}$\,cm$^{-2}$ (Wang et al., submitted). 
Thirteen \Oiii\, sources are matched to 10 LBGs with $6<z_{phot}<6.8$ in this work (IDs 267 and 823 are resolved into three and two components respectively but were included in one Kron aperture in this work), all of which agree spectroscopically within 1$\sigma$ of the best fit EAZY redshift (2$\sigma$ for the aforementioned blended sources).
The remaining \Oiii\, emitters were not classified as LBGs at the correct redshift, either having $z_{phot} < 5$ or lying just outside our required $z_{qso} \pm 0.7$.
The high rate of overlap between these independently derived catalogs validates the robustness of our selection criteria, especially considering that four of these matches were classified as ``marginal" in our sample.
A future paper will be dedicated to further characterizing the properties of the spectroscopically confirmed overdensity members (Champagne et al., in prep.).

We note that 35/42 of the sources detected within $\Delta z = 0.7$ from J0305 QSO had best-fitting redshifts in front of the quasar, with a median $z_{phot}=6.24$.
This includes the eastern ``filament," which contains 16 galaxies with photo-$z$'s between 6.0 and 6.2.
To see if this remained significant at lower redshifts, we repeated the search for LBGs centered on $z<6.6$ iteratively in steps of $z=0.1$.
We find that the strongest overdensity signal is when we center the search on $z=6.2$ where the median $z_{phot} = 6.16$.
Here, we find a total of 49 LBGs, with 8 sources being unique from our reported J0305 QSO sample (three of which are also reported as \Oiii\, emitters in Wang et al. submitted), yielding an overdensity $\delta_{\rm gal} = 10.9 \pm 2.0$.
Given the uncertainties in photometric redshifts\footnote{See Appendix \ref{sec:templates} for a discussion of how our SED templates affect the best-fit redshift, since our assumption that Ly$\alpha$ is always attenuated results in systematically lower-redshift $z_{\rm phot}$ distributions than when we include Ly$\alpha$ with 100\% escape.}, the strong 2D clustering of this filament could indicate association with the quasar or a separate overdensity along the line of sight. 
Separately, the parallel field is also marginally overdense at the 3$\sigma$ level.
Given that the J0305 overdensity extends across the full field of view of WFC3, an additional overdensity separated by a few Mpc could also potentially lend credence to the existence of a protocluster spanning more than 10 Mpc in the transverse direction.

We next evaluate the probability of observing \textit{N} galaxies in a given redshift interval \textit{dz} following the methodology of \citet{Steidel1998a}, who characterized the significance of the LBG overdensity in the SSA22 field using the Erlang distribution. 
The probability of observing \textit{N} galaxies in the absence of clustering is given by:

\begin{equation}
    p(\Delta z | N \lambda) = \lambda (\lambda\Delta z)^{N-2} \exp{(-\lambda\Delta z)} / (N-2)!
\end{equation}

Here, $\lambda$ is the expected number of galaxies per unit redshift interval which can be obtained by multiplying the differential volume element by a selection function, in this case the UVLF from \citet{Finkelstein2016a} evaluated at $z=6.6$ down to $M_{UV} < -19.5$. 
The volume sampled by the total \textit{secure} $\Sigma P(z)$ is $\Delta z = 0.9$, in which we would expect $\lambda = 4$ galaxies within the full WFC3 field of view.
WFC3 spans $\sim 2.5$ comoving Mpc on the sky, which is much smaller than the expected physical span of a $z\sim7$ protocluster of 10--20 comoving Mpc \citep{Chiang2017a}.
The probability of randomly observing 18 LBGs with no clustering, in the same comoving volume centered on the quasar, is 0.004\%.
If we include all of the secure LBGs plus an assumption that 50\% of the marginal LBGs lie at their best-fit redshifts (the median weight applied in $\S$\ref{sec:uvlf}), the chance of observing them randomly is $\sim10^{-11}$.
Therefore, even with our relatively uncertain volume constraints across $\Delta z = 1$, there is strong evidence to support that the 2D overdensity is not simply due to a chance projection along the line of sight.

\begin{figure}
    \centering
   \includegraphics[trim={0.1in 0.05in 0.1in 0.05in}, width=1.0\columnwidth]{ 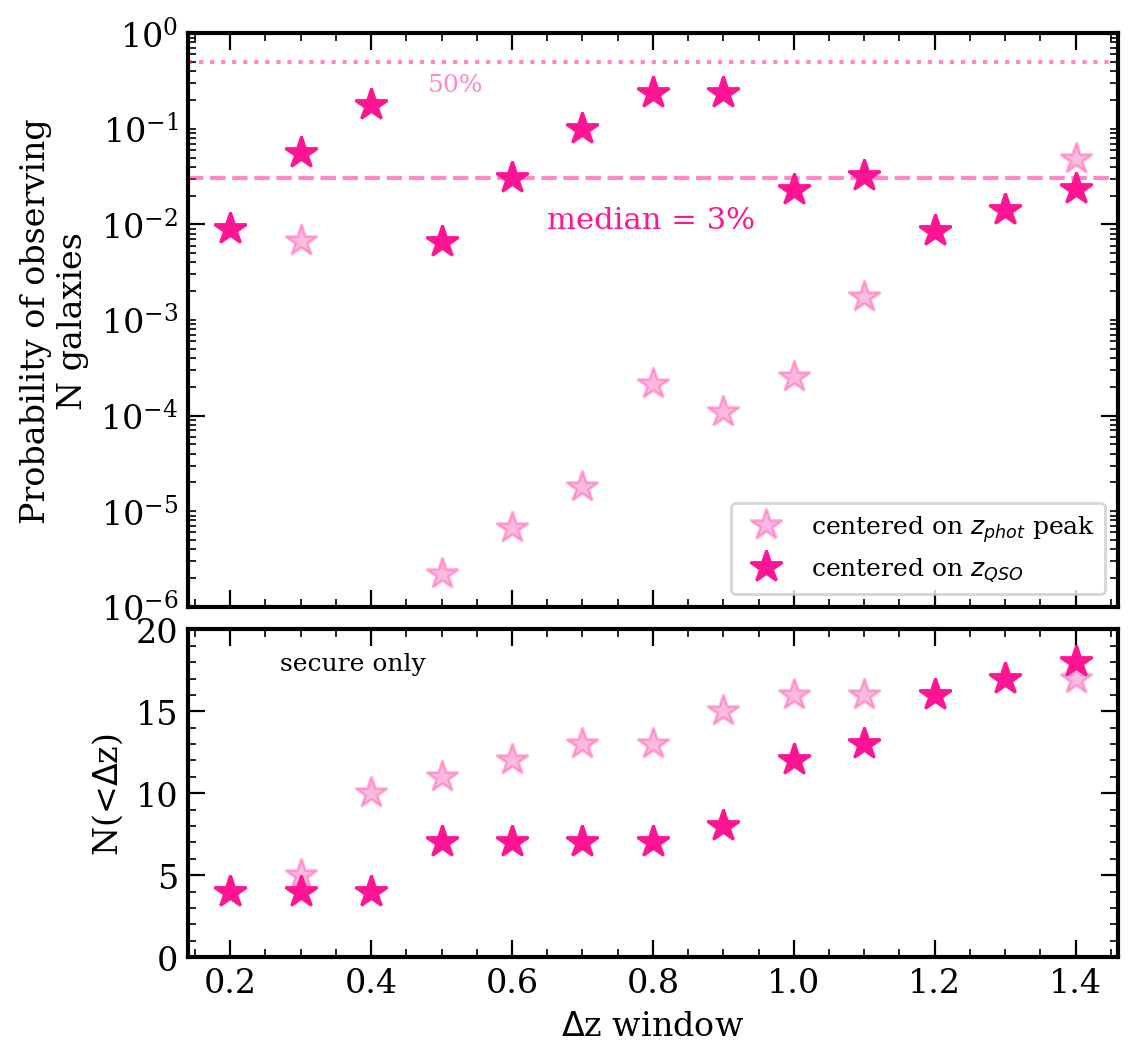}
    \caption{The probability of observing LBGs in the J0305 field by chance, i.e. in the absence of true clustering, as a function of $\Delta z$. The top panel shows the $P(\Delta z|N\lambda)$ statistic and the bottom panel shows the number of \textit{secure} galaxies with best-fit redshifts in the $\Delta z$ window. The dark pink stars mark the secure galaxies in this sample with the statistic centered on the quasar at $z=6.6$, while the lighter pink stars are evaluated at the peak $z_{phot} = 6.2$. 
    Even though the probability of a chance alignment of galaxies is much less than 1\% in a reasonably large $\Delta z$ window of 0.5, the overdensity signal gets washed out when $\Delta z$ is large enough to encompass the background density.  
    \label{fig:pdzlam}
    }
\end{figure}

A more physically motivated scale in which to evaluate the probability of a protocluster structure would be closer to $\Delta z \sim 0.1$.
We can reasonably apply this statistic to smaller physical scales by assuming our photo-$z$'s for the secure candidates are correct and iteratively calculating the probability of observing $N$ LBGs as a function of $dz$.
Figure \ref{fig:pdzlam} shows the probability of observing \textit{N} LBGs in the J0305 field.
Between $\Delta z = 0.2$ and $\Delta z = 1$, the median probability of observing $N$ = 4--18 LBGs in the absence of clustering is 3\% with an rms spread of 0.015\% centered on the peak $z_{phot}$, indicating that the overdensity is clear regardless of the redshift interval we choose. 
Importantly, this figure shows that increasingly wider $\Delta z$ assumptions begin to wash out the rare overdensity beyond $\Delta z \sim 0.5$.

Finally, we examine the angular distribution of the LBGs in J0305, noting an avoidance of UV-bright sources at close angular separations.
All of the sources within 10\arcsec\, of the quasar are fainter than $M_{UV} = -20.5$; however, all three of them are in the marginal sample so do not consider their positions constrained to be near the quasar.
There are 4 secure sources with $M_{UV} < -20.7$ and $z_{phot}$ within $z_{qso}\pm0.1$, which are all separated by more than 20\arcsec\, (120 proper kpc) but less than 50\arcsec.
For reference, the volume density of $M_{UV} < -21$ sources is expected to be $<4 \times 10^{-5}$\,Mpc$^{-3}$, so even pessimistically we would expect $<0.5$ bright sources within $\Delta z = 2$, or $< 0.05$ bright sources within $\Delta z = 0.2$.
Thus, there is an order of magnitude overdensity of the brightest sources, but only at large separations.

On the other hand, \citet{Farina2017a} found one faint LAE companion within the arcmin field of view of MUSE, separated by 12 kpc (2\arcsec) and assumed to be undergoing a merger with the quasar host galaxy; this companion is detected with SNR = 2 in \hband\, in this data, thus it was not included in our LBG list.
Similarly, the overdensity reported by Wang et al. (submitted) shows an enhancement of bright \Oiii\, emitters with $f_{\Oiii} > 5.0 \times 10^{-18}$\,erg\,s$^{-1}$\,cm$^{-2}$ within 10\arcsec\, from the quasar, which had faint but visible counterparts in the \textit{HST} data that did not meet our \hband\, detection criteria. 
In a Subaru Suprime-Cam survey in the J0305 field, \citet{Ota2018a} also noted a dearth of faint LAEs and LBGs given their 50\% completeness limit of $m = 26$.
We thus conclude that the overdensity very close to the quasar is primarily made up of relatively fainter galaxies, but the strongest overdensity of bright galaxies exists in the filament approximately 500 transverse kpc away at $z\approx6.2$. 

\citet{Bosman2020a} discusses the visibility of LAEs within the proximity zone of $z\sim6$ quasars, noting the possible suppression effect of ionizing radiation from the quasar within a few proper Mpc and  concluding that lower-mass galaxies would be preferentially suppressed during the quasar's active phase.
However, LBGs with $M_{UV} < -19.5$, like the ones probed by our data, occupy larger halos of $M_h > 10^{11}$\,\msun\, \citep{Harikane2018a} and are not expected to be strongly suppressed.
\citet{Lidz2006a} instead suggests the idea that while quasars are \textit{born} in overdense regions of the Universe, once the quasar turns on it outshines the surrounding galaxies and dominates the local photoionization rate during the quasar lifetime.
\citet{Utsumi2010a} suggest, along the same lines, that quasar activity is triggered in regions where luminous, massive galaxies have already evolved.
Thus, by the time we observe a luminous quasar, the galaxies in its immediate environment may have already shut down star formation or merged with the host galaxy, therefore we would instead observe \textit{under}-densities in its immediate environment, and \textit{over}-densities at further separations.
This ``ring" is suggested by both hydrodynamical simulations \citep[e.g.,][]{Habouzit2019a} and observations of $z\sim5-6$ quasars \citep{Utsumi2010a}, although on 10$\times$ wider scales than our results --- comparable to the expected size of the proximity zone of $\sim 3$ pMpc \citep{Eilers2017a}.

\subsubsection{The nature of the J2054 overdensity}\label{sec:j2054}

J2054 is the most UV-luminous quasar in our sample, observed close to the end of the EoR.
In other studies, this quasar field showed only one \Cii\, emitter 77 kpc away \citep{Venemans2020a} and a dust continuum emitter separated by 60 transverse kpc but without a redshift \citep{Champagne2018a}, so no previous long-wavelength studies have suggested an overdensity. 
We noted earlier that \citet{Marshall2020a} reported several candidate companion galaxies $<10$\arcsec\, from the quasar as well, selected in \jband\, and \hband.
While we do not find any correlation between separation from the quasar and LBG UV luminosity for J2054, we note that the two sources within 10\arcsec\, ($\sim60$\,kpc) of the quasar have M$_{UV} < -21.5$, one of which has a $z_{phot} = 5.95$.
Across the full field of view, all members within $\Delta z = 0.1$ from the quasar have M$_{UV} < -21$.
Opposite to J0305, we see a strongly centrally peaked distribution of very bright galaxies within 1 cMpc as noted in Figure \ref{fig:dgalr}. 
In fact, using the Horizon-AGN simulations \citet{Habouzit2019a} noted that the strongest overdensity signal would indeed be measured at $<$1 cMpc while the radial distribution would resemble the field at larger separations.  
Thus, while the 3D number counts across the full field of view are only modestly overdense at 2.7$\sigma$, we argue that the $\sim25\times$ angular overdensity at close separations constitutes a statistically significant overdensity even if the LBGs lie at further separations along the line of sight.
Using the $p(\Delta z|N\lambda)$ statistic, we find there is a 10\% chance of observing the galaxies within 30\arcsec\, by chance. 
Such close-in overdensities have also been noted in other photometric quasar environment studies at $z\sim5-6$ \citep[e.g.,][]{Kim2009a, McGreer2014a, Trakhtenbrot2017a}, some of which have been spectroscopically confirmed \citep[e.g.,][]{Zheng2006a, Bosman2020a, Overzier2022a}.

\subsubsection{The nature of the J2348 environment}
Finally, J2348 lies at the highest redshift and contains the lowest density of galaxies in its vicinity.
The 2D number counts are completely consistent with the blank field, but the quasar lies 2$\sigma$ above the median of $\Sigma P(z)$, lending some credence to a local underdensity (though not spectroscopically confirmed). 
Recent high-resolution imaging of the J2348 host galaxy revealed unusually compact dust and \Cii\, emission \citep[1 kpc in diameter;][]{Walter2022a} in the central region of the host galaxy, so feedback is an unlikely reason for the low density of galaxies.
It has an extremely high SFR surface density of $10^4$\,\msun\,yr$^{-1}$\,kpc$^{-2}$ and contains the most massive black hole in our sample, yet we do not see an enhancement of the surrounding environment.
\citet{Champagne2018a} showed a dust continuum emitter separated by 41 transverse kpc, while no previous \Cii\, studies \citep{Decarli2018a, Venemans2020a} reported any companions.

While one might naively expect the most massive SMBH in the sample to host the strongest overdensities, \citet{Fanidakis2013a} suggested that the most massive black holes may not occupy the most massive dark matter halos but instead form among multiple smaller halos, meaning that the galaxy number counts do not correlate with SMBH mass. 
\citet{Habouzit2019a} also simulate SMBHs that form in less extreme dark matter overdensities, where real galaxy overdensities might only be found along filaments substantially separated from the quasar.
It is still entirely possible that overdensities would be revealed in wider-field observations, motivating current and upcoming quasar surveys taking advantage of wider-field mosaics \citep[e.g., EIGER, which has already revealed a marked overdensity on $\sim0.1-2$ arcmin scales around a $z=6.3$ quasar using slitless spectroscopy][]{Kashino2022a},
or that ionizing radiation within an isotropic proximity zone of $\sim3-5$\, pMpc would prevent a detectable overdensity signal on larger scales.
Still, we stress that a $M>10^9$\,\msun\, SMBH does not guarantee an overdensity of galaxies exists nearby, which has also been supported in prior overdensity searches yielding underdensities or no enhancement around $z\sim7$ quasars \citep[e.g.,][]{Simpson2014a}.

\subsection{Comparison to Previous ALMA Observations}\label{sec:dsfg}
We now compare the LBG results with the 1.1\,mm sources detected by ALMA, probing the dust-obscured galaxy population.
J0305 and J2054 each contain one \Cii\, companion within 100 kpc, and all three quasars have at least one dust continuum emitter within the 20\arcsec\, ALMA primary beam.
None of the LBGs in this sample are coincident with a millimeter continuum detection, suggesting that a) the existing dust continuum sources may not lie anywhere near the quasar redshift, and b) the LBGs presented here are not significantly attenuated by dust (as expected since they all show very blue UV continuum slopes).
Thus, in the immediate vicinity of the quasars, there is no detection of a preferential enhancement of very dusty star formation that frequently accompanies protocluster structures at $z>4$ \citep[e.g.,][]{Harikane2019a, Hill2020a, Long2020a}.
However, the ALMA data probes only the inner few hundred kpc near the quasar, and as we noted in $\S$\ref{sec:uvlf}, J2054 and J0305 both show markedly stronger overdensities of bright LBGs at wider separations; thus, ALMA mosaics on arcmin scales would be required to make a fair comparison of the IR-bright versus the UV-bright populations.

\subsection{Accelerated Growth of LBGs?}\label{sec:acc}
In the J0305 and J2054 quasar fields, we have noted an apparently stronger overdensity signal at the massive end of the UVLF.
Quantitatively, 30\% of the LBGs in J0305 have $M<M^{*}_{z=6.6}$ and 60\% in J2054 have $M<M^*_{z=6}$; even without correcting for completeness, this is significantly above the expected fraction of $M<M^* = 10$\% at the flux limit of our observations.
We can compare our results with luminosity functions at other redshifts to qualitatively imply accelerated evolution of the UV luminous population, since predictions from the evolution of the UVLF imply that massive galaxies become more abundant and finish growing at earlier times.
To demonstrate accelerated growth in these quasar fields, we compare our LF results with literature LFs at $z=4-7$ \citep{Finkelstein2015a}.
We can see from Figure \ref{fig:uvlf} that the slope of the bright end in both J0305 and J2054 is most consistent with the $z=4$ UVLF, approximately 600 Myr later. 
Indeed, this points to the fact that more massive sub-halos in overdense regions are able to grow at an accelerated pace compared to random fields, such that the galaxies reach their most luminous and massive state earlier than the average evolution of the UVLF at the same redshift.
Given the detection of several \Oiii\, emitters close to J0305 in addition to the nearby LAE, neither of which have strong rest-UV counterparts, the central overdensity may be in nascent stages of merger activity and accelerated star formation, while the outer/lower-redshift overdensities display evidence of having already experienced rapid formation. 

\subsection{Comparisons to Model Predictions: Implications for Detecting Overdensities}\label{sec:models}
Finally, semi-analytical models have predicted it would be difficult to measure the amplitude of a protocluster-scale overdensity at $z=6$, where the dark matter overdensity $\delta_m$ is much smaller than in protoclusters at lower redshifts ($z<3$) that are nearing virial collapse.
\citet{Chiang2013a} showed using N-body simulations that even the most massive present-day clusters with $M_h \sim 10^{15}$\,\msun\, would have a predicted $\delta_{\rm gal} \sim 8$ at $z\sim5$.
The authors also show that the predicted 3D $\delta_{\rm gal}$ in finite boxes of ($\sim10-25$ Mpc)$^3$ does not take into account the limits of real observations --- as $\Delta z$ increases, the density field is smoothed and the observed $\delta_{\rm gal}$ quickly resembles the field, even for truly clustered populations.
Thus, one can only pick up the most overdense systems with a photo-$z$ study, since the observed $\delta_{\rm gal}$ would need to be much larger to overcome the diminished signal induced by a wide $\Delta z$. 
Additionally, because the 2D overdensity profile of a theoretical protocluster is strongly centrally peaked and flattens to $\delta_{\rm gal}\sim0$ within a few comoving Mpc \citep{Chiang2013a, Chiang2017a}, the projection of members on the outskirts will also dilute a real overdensity signal.
One can see this effect in Figure \ref{fig:dgalr}, where J2054 shows a projected $\delta_{gal,2D} \sim 25$ within 500 comoving kpc from the quasar but is indistinguishable from the field at $R>1$\,Mpc, which is indeed the predicted overdensity profile for a $z=5$ progenitor of a Coma-like cluster in \citet{Chiang2013a}. 
In contrast, J0305 shows a fairly flat 2D $\delta_{\rm gal}$ of $\sim10$ out to the edge of the WFC3 field of view, implying that we are sensitive to an extended overdensity regardless of the physical limits we consider.

In the same vein, a diversity of environments is expected around quasars at $z>5$, as is discussed in detail in \citet{Habouzit2019a} who used hydrodynamical simulations to predict the galaxy number counts around quasars.
While AGN feedback can suppress star formation in massive galaxies, this is expected to be a smaller effect than the anisotropy of the quasar's surrounding H\textsc{ii} region, giving rise to different expected number counts of 0--10 galaxies within 1 comoving Mpc of the quasar. 
Finally, they note that the definition of an overdensity depends heavily on the classification of galaxy used to probe it --- for example, setting a high stellar mass or luminosity threshold will dilute the amplitude of an overdensity for an intrinsically rarer population of galaxies. 
Thus, probing fainter ($M_{UV} > -19$) galaxies with \textit{JWST} and across wider scales that encompass the full extent of filamentary overdensities (with, e.g, \textit{Euclid}) is the next frontier in exploring quasar environments. 

\section{Conclusions}\label{sec:conclusion}
In this work, we presented a search for LBGs in the vicinities of three luminous quasars at $6<z<7$ using 5 filters of \textit{HST} broadband imaging. 
We identified dropout candidates and constrained their photometric redshifts using EAZY and presented a series of methods to improve the robustness of our selection criteria.
Using only the 2D positional information, we compared the spatial distribution of LBGs with expectations from the blank field UVLF.
We also measured the effective volume spanned by the LBGs by co-adding the  \pz\, distributions of individual candidates and calculated the UVLF in bins of $\Delta z \approx 1$ around the quasars.
The results show a range of overdensity signals in two- and three-dimensional analyses, most notably a stronger signal when considering the brightest galaxies in the sample.
Finally, we discussed a number of physical scenarios that would give rise to the observed overdensity signals, including serendipitous overdensities along the line of sight, limitations in photo-$z$ constraints, and the genuine diversity of quasar environments expected due to, e.g., ionizing radiation within the proximity zone.
Below we summarize the main results from this study:

\begin{itemize}
\item J0305--3150, at $z=6.6$, hosts the strongest overdensity in the sample with 42 LBGs, corresponding to $\delta_{\rm gal} = 8.8 \pm 1.8$. 
In a companion paper, Wang et al. (submitted) shows a spectroscopically confirmed overdensity of \Oiii\, emitters, 13 of which correspond to LBGs reported in this paper. 
Its coordinated parallel field shows a significant 3$\sigma$ overdensity, which could be consistent with being part of the same structure.
At short angular separations ($<$12\arcsec\, or 0.5 comoving Mpc), the two-dimensional overdensity rises to $\delta_{gal,2D} > 30$, and we see evidence for a separate filamentary structure within $z_{phot} = 6.0-6.2$ approximately 30\arcsec\, from the quasar.

\item J2054--0005 shows a marked 2D overdensity within 0.5 cMpc aperture, but drops to blank field expectations at wider separations.
In 3D, it shows a $\sim$3$\sigma$ overdensity of $\delta_{\rm gal} = 1.9 \pm 0.7$, with 4 LBGs with best-fit redshifts within $\Delta z = 0.1$ from the quasar and 18 LBGs total.
The parallel field is consistent with the blank field with $\delta_{\rm gal} = 0.1\pm 0.4$.

\item J2348--3054 at $z=6.9$ shows the lowest density of sources, with no enhancement of LBGs in the primary and parallel fields in either 2D or 3D. 
With 6 LBGs in the primary and 4 in the parallel, they show $\delta_{\rm gal} = 1.3 \pm 1.2$ and $0.0 \pm 0.6$ respectively.

\item We discuss the spatial distribution of the overdensities, noting that the overdensity in J2054 is strongly centrally peaked but more uniform in J0305. 
We hypothesize on the effects of ionizing radiation from the quasar in suppressing faint companions as well as the possibility that accelerated evolution in overdense regions of the cosmic web can result in a strong overdensity of brighter galaxies.

\item We discuss the difficulty in assessing a 3D overdensity in small fields of view which is expected to be modest at high redshifts, as supported by simulations. 
Too large an assumed cosmic volume can wash out a true overdensity, but a genuine diversity of quasar environments is expected from hydrodyamical simulations.
Similarly, comparing to the average 2D number counts can also wash out an overdensity at large angular separations, but this is expected based on the fact that the overdensity profiles are predicted to be strongly centrally peaked but modest at larger radii.

\end{itemize}

Though a sample of three quasar fields does not represent a statistically representative sample, we have shown heterogeneity across three different fields between $6<z<7$, in line with previous studies showing a variety of quasar environments that do not correlate with any of the physical properties of the quasar or its host galaxy. 
Spectroscopic confirmation is needed to strengthen the claim that J2054 presents a modest overdensity, and wider-field observations are necessary to strengthen the idea that quasars may not lie at the virial centers of bona fide overdensities and are thus missed by small-scale observations.
We have demonstrated potentially accelerated galaxy evolution particularly in the J0305 field, which is investigated further in Wang et al., (submitted). 
We have stressed that different definitions of the overdensity $\delta_{\rm gal}$ yield significantly different results and have attempted to mitigate these effects through robust LBG selection and physically motivated measurements.

Given the relatively short lifetime of the quasar's active phase, it is likely they are not ubiquitous tracers of overdense environments at all times.
However, the strong preferential enhancement of massive galaxies in two out of the three quasar fields implies that the overdense environment effects rapid galaxy evolution in the vicinity of a quasar.
Overall, we have shown that a search for LBGs even in single \textit{HST} pointings can be effective in detecting the most significant overdensities and offer insight into the diversity of quasar environments during the EoR. 
Future surveys with \textit{JWST} and \textit{Euclid} will allow us to map wider fields and gain spectroscopic confirmation of the hints at overdensities we have seen thus far.

\begin{deluxetable*}{ccccccccc}
\tablecaption{\textbf{J0305 QSO field.} LBG candidate positions with SE catalog ID number, best fit redshift from EAZY, and measured \textit{HST} photometry. Fluxes are given in units of $1.0\times10^{-8}$ Jy. \label{table:j0305q}}
\tablehead{\colhead{ID} & \colhead{RA} & \colhead{Dec} & \colhead{photo-$z$} & \colhead{F606W} & \colhead{F814W} & \colhead{F105W} & \colhead{F125W} & \colhead{F160W}}
\startdata
\sidehead{Secure}
267 & 03:05:19.9 & $-$31:50:19.6 & 6.8 $\pm$ 0.5 & 0.4 $\pm$ 1.1 & 1.9 $\pm$ 1.4 & 22.9 $\pm$ 1.6 & 25.3 $\pm$ 1.2 & 18.8 $\pm$ 1.2 \\
4945 & 03:05:11.0 & $-$31:51:37.0 & 6.5 $\pm$ 0.3 & 0.5 $\pm$ 1.3 & 3.8 $\pm$ 1.9 & 12.1 $\pm$ 1.9 & 7.2 $\pm$ 1.4 & 6.2 $\pm$ 1.4 \\
3057 & 03:05:15.9 & $-$31:51:16.9 & 6.5 $\pm$ 0.5 & 3.5 $\pm$ 1.0 & 1.9 $\pm$ 1.4 & 16.2 $\pm$ 1.4 & 11.9 $\pm$ 1.1 & 9.8 $\pm$ 1.1 \\
2009 & 03:05:20.2 & $-$31:51:43.0 & 6.1 $\pm$ 0.4 & 1.3 $\pm$ 1.1 & 5.2 $\pm$ 1.5 & 21.3 $\pm$ 1.5 & 22.6 $\pm$ 1.2 & 22.3 $\pm$ 1.2 \\
2146 & 03:05:15.7 & $-$31:50:40.8 & 7.1 $\pm$ 0.5 & $-$1.1 $\pm$ 1.5 & $-$2.6 $\pm$ 2.1 & 16.9 $\pm$ 2.2 & 11.5 $\pm$ 1.7 & 11.0 $\pm$ 1.7 \\
4350 & 03:05:14.5 & $-$31:51:47.1 & 6.5 $\pm$ 0.3 & $-$0.4 $\pm$ 1.3 & 4.5 $\pm$ 1.8 & 14.9 $\pm$ 1.9 & 10.1 $\pm$ 1.4 & 11.2 $\pm$ 1.4 \\
4152 & 03:05:13.0 & $-$31:51:14.9 & 6.6 $\pm$ 0.6 & $-$1.7 $\pm$ 1.0 & $-$0.3 $\pm$ 1.4 & 14.2 $\pm$ 1.4 & 12.1 $\pm$ 1.1 & 8.0 $\pm$ 1.1 \\
1552 & 03:05:19.3 & $-$31:51:13.7 & 6.1 $\pm$ 0.4 & 1.8 $\pm$ 1.1 & 6.8 $\pm$ 1.4 & 27.8 $\pm$ 1.5 & 23.0 $\pm$ 1.2 & 21.0 $\pm$ 1.2 \\
1457 & 03:05:19.1 & $-$31:51:08.6 & 6.1 $\pm$ 0.6 & 0.4 $\pm$ 0.8 & 2.5 $\pm$ 1.1 & 9.9 $\pm$ 1.1 & 8.9 $\pm$ 0.9 & 8.1 $\pm$ 0.9 \\
2193 & 03:05:18.0 & $-$31:51:19.0 & 5.9 $\pm$ 0.4 & 1.4 $\pm$ 1.2 & 6.4 $\pm$ 1.7 & 20.3 $\pm$ 1.9 & 18.5 $\pm$ 1.5 & 15.9 $\pm$ 1.5 \\
470 & 03:05:19.7 & $-$31:50:30.0 & 6.5 $\pm$ 0.1 & 0.6 $\pm$ 1.2 & 6.8 $\pm$ 1.6 & 26.1 $\pm$ 1.7 & 21.0 $\pm$ 1.3 & 18.8 $\pm$ 1.3 \\
1456 & 03:05:19.1 & $-$31:51:08.0 & 6.5 $\pm$ 0.3 & 0.2 $\pm$ 0.7 & 2.3 $\pm$ 0.9 & 8.0 $\pm$ 1.0 & 5.2 $\pm$ 0.8 & 5.3 $\pm$ 0.8 \\
961 & 03:05:19.7 & $-$31:50:59.1 & 6.5 $\pm$ 1.0 & 0.5 $\pm$ 1.1 & 0.8 $\pm$ 1.5 & 9.8 $\pm$ 1.6 & 8.3 $\pm$ 1.2 & 6.4 $\pm$ 1.2 \\
1273 & 03:05:18.8 & $-$31:50:56.0 & 7.1 $\pm$ 0.6 & 1.3 $\pm$ 1.3 & $-$2.0 $\pm$ 1.7 & 13.6 $\pm$ 2.1 & 8.3 $\pm$ 1.5 & 9.2 $\pm$ 1.5 \\
3232 & 03:05:16.7 & $-$31:51:36.0 & 7.1 $\pm$ 0.8 & 3.1 $\pm$ 1.3 & $-$0.9 $\pm$ 1.9 & 12.9 $\pm$ 1.8 & 15.1 $\pm$ 1.4 & 11.6 $\pm$ 1.4 \\
1366 & 03:05:15.6 & $-$31:50:13.3 & 6.4 $\pm$ 0.5 & 3.5 $\pm$ 2.3 & 2.6 $\pm$ 2.8 & 19.6 $\pm$ 2.5 & 17.1 $\pm$ 1.9 & 9.9 $\pm$ 1.9 \\
589 & 03:05:20.3 & $-$31:50:49.5 & 6.0 $\pm$ 0.4 & 1.5 $\pm$ 1.1 & 5.0 $\pm$ 1.7 & 17.2 $\pm$ 1.7 & 14.5 $\pm$ 1.3 & 10.0 $\pm$ 1.3 \\
1468 & 03:05:19.8 & $-$31:51:19.8 & 6.5 $\pm$ 0.3 & 0.8 $\pm$ 1.1 & 3.9 $\pm$ 1.6 & 18.4 $\pm$ 1.6 & 13.8 $\pm$ 1.2 & 12.9 $\pm$ 1.2 \\
\sidehead{Marginal}
3638 & 03:05:16.3 & $-$31:51:43.8 & 6.8 $\pm$ 0.6 & 0.9 $\pm$ 1.1 & 0.7 $\pm$ 1.6 & 9.9 $\pm$ 1.7 & 4.1 $\pm$ 1.3 & 5.8 $\pm$ 1.3 \\
366 & 03:05:20.3 & $-$31:50:34.2 & 6.3 $\pm$ 1.7 & $-$2.0 $\pm$ 1.3 & 1.8 $\pm$ 2.0 & 5.9 $\pm$ 1.9 & 6.7 $\pm$ 1.4 & 4.3 $\pm$ 1.4 \\
3251 & 03:05:16.9 & $-$31:51:32.0 & 6.9 $\pm$ 0.7 & 0.3 $\pm$ 1.1 & 0.2 $\pm$ 1.4 & 6.9 $\pm$ 1.5 & 5.8 $\pm$ 1.2 & 4.9 $\pm$ 1.2 \\
2109 & 03:05:19.7 & $-$31:51:38.8 & 7.1 $\pm$ 0.6 & $-$2.9 $\pm$ 1.3 & $-$0.2 $\pm$ 1.8 & 12.1 $\pm$ 1.8 & 9.4 $\pm$ 1.4 & 11.4 $\pm$ 1.4 \\
28 & 03:05:22.2 & $-$31:50:14.6 & 6.9 $\pm$ 0.7 & 0.2 $\pm$ 1.5 & 1.1 $\pm$ 2.0 & 14.3 $\pm$ 3.6 & 6.6 $\pm$ 3.0 & 8.6 $\pm$ 2.5 \\
1201 & 03:05:15.3 & $-$31:50:03.0 & 6.5 $\pm$ 0.1 & 0.4 $\pm$ 1.1 & 5.8 $\pm$ 1.6 & 14.9 $\pm$ 1.6 & 8.0 $\pm$ 1.2 & 5.4 $\pm$ 1.2 \\
2179 & 03:05:19.4 & $-$31:51:37.2 & 6.8 $\pm$ 0.7 & 0.3 $\pm$ 1.1 & 1.3 $\pm$ 1.5 & 9.5 $\pm$ 1.6 & 7.1 $\pm$ 1.3 & 7.1 $\pm$ 1.3 \\
1116 & 03:05:01.0 & $-$31:49:48.2 & 6.6 $\pm$ 0.7 & $-$1.4 $\pm$ 1.0 & 1.7 $\pm$ 1.2 & 7.8 $\pm$ 1.2 & 5.7 $\pm$ 1.0 & 7.2 $\pm$ 1.0 \\
2335 & 03:05:14.0 & $-$31:50:21.1 & 6.4 $\pm$ 0.5 & $-$1.0 $\pm$ 1.2 & 3.0 $\pm$ 1.6 & 10.3 $\pm$ 1.7 & 9.6 $\pm$ 1.3 & 8.4 $\pm$ 1.3 \\
3069 & 03:05:19.6 & $-$31:52:13.0 & 6.5 $\pm$ 0.4 & 0.5 $\pm$ 1.0 & 2.1 $\pm$ 1.4 & 7.1 $\pm$ 1.5 & 5.0 $\pm$ 1.2 & 4.3 $\pm$ 1.2 \\
2166 & 3:05:16.0 & $-$31:50:51.0 & 6.3 $\pm$ 1.6 & $-$0.9 $\pm$ 1.5 & 1.2 $\pm$ 1.9 & 7.6 $\pm$ 2.1 & 7.7 $\pm$ 1.6 & 5.4 $\pm$ 1.6 \\
823 & 03:05:18.7 & $-$31:50:38.4 & 6.9 $\pm$ 0.8 & 0.3 $\pm$ 0.9 & 0.7 $\pm$ 1.2 & 13.1 $\pm$ 1.4 & 14.3 $\pm$ 1.1 & 15.2 $\pm$ 1.1 \\
1676 & 03:05:18.0 & $-$31:50:59.0 & 6.9 $\pm$ 0.7 & 2.0 $\pm$ 1.2 & 0.8 $\pm$ 1.6 & 9.5 $\pm$ 1.8 & 8.2 $\pm$ 1.4 & 7.6 $\pm$ 1.4 
\enddata
\end{deluxetable*}

\begin{deluxetable*}{ccccccccc}
\tablecaption{Table \ref{table:j0305q} continued.}
\tablehead{\colhead{ID} & \colhead{RA} & \colhead{Dec} & \colhead{photo-$z$} & \colhead{F606W} & \colhead{F814W} & \colhead{F105W} & \colhead{F125W} & \colhead{F160W}}
\startdata
25 & 03:05:22.2 & $-$31:50:14.0 & 6.3 $\pm$ 1.3 & $-$0.0 $\pm$ 1.0 & 1.8 $\pm$ 1.5 & 6.7 $\pm$ 1.8 & 6.9 $\pm$ 1.3 & 3.8 $\pm$ 1.4 \\
4698 & 03:05:11.4 & $-$31:51:18.7 & 6.8 $\pm$ 0.7 & $-$0.4 $\pm$ 1.5 & 1.2 $\pm$ 2.0 & 11.1 $\pm$ 2.1 & 5.3 $\pm$ 1.6 & 7.8 $\pm$ 1.6 \\
1813 & 03:05:18.0 & $-$31:51:03.9 & 6.0 $\pm$ 0.9 & 1.3 $\pm$ 1.7 & 3.3 $\pm$ 2.4 & 10.4 $\pm$ 2.4 & 8.6 $\pm$ 1.8 & 6.8 $\pm$ 1.8 \\
1010 & 03:05:18.0 & $-$31:50:44.7 & 6.8 $\pm$ 0.7 & $-$1.8 $\pm$ 1.3 & 1.2 $\pm$ 1.8 & 8.0 $\pm$ 1.9 & 5.4 $\pm$ 1.4 & 7.1 $\pm$ 1.4 \\
1041 & 03:05:20.2 & $-$31:51:00.0 & 6.3 $\pm$ 1.5 & 0.1 $\pm$ 1.2 & 1.4 $\pm$ 1.6 & 5.9 $\pm$ 1.7 & 6.0 $\pm$ 1.3 & 3.0 $\pm$ 1.3 \\
4553 & 03:05:13.0 & $-$31:51:32.0 & 5.9 $\pm$ 0.8 & 0.4 $\pm$ 1.2 & 2.9 $\pm$ 1.6 & 8.2 $\pm$ 1.7 & 6.3 $\pm$ 1.3 & 5.5 $\pm$ 1.3 \\
4452 & 03:05:15.0 & $-$31:51:58.0 & 7.1 $\pm$ 0.6 & $-$1.6 $\pm$ 1.0 & $-$1.1 $\pm$ 1.3 & 6.1 $\pm$ 1.4 & 3.5 $\pm$ 1.1 & 4.2 $\pm$ 1.1 \\
351 & 03:05:18.7 & $-$31:50:07.7 & 6.4 $\pm$ 1.0 & $-$0.3 $\pm$ 0.8 & 1.5 $\pm$ 1.1 & 6.1 $\pm$ 1.2 & 5.7 $\pm$ 1.0 & 4.8 $\pm$ 0.9 \\
4783 & 03:05:11.0 & $-$31:51:02.0 & 6.5 $\pm$ 0.6 & $-$0.3 $\pm$ 1.6 & 2.7 $\pm$ 2.2 & 12.1 $\pm$ 2.3 & 6.3 $\pm$ 1.8 & 6.3 $\pm$ 1.8 \\
2442 & 03:05:16.6 & $-$31:51:02.2 & 7.5 $\pm$ 0.2 & $-$1.7 $\pm$ 1.0 & $-$0.5 $\pm$ 1.5 & 5.4 $\pm$ 1.4 & 3.9 $\pm$ 1.1 & 5.7 $\pm$ 1.0 \\
194 & 03:05:19.4 & $-$31:50:03.7 & 7.0 $\pm$ 0.7 & 1.1 $\pm$ 1.6 & 0.3 $\pm$ 2.2 & 10.0 $\pm$ 2.6 & 7.4 $\pm$ 2.0 & 7.5 $\pm$ 2.0 \\
2437 & 03:05:16.2 & $-$31:50:59.0 & 6.0 $\pm$ 1.0 & $-$1.1 $\pm$ 1.6 & 3.5 $\pm$ 2.3 & 9.1 $\pm$ 2.2 & 6.7 $\pm$ 1.7 & 4.9 $\pm$ 1.7 
\enddata
\end{deluxetable*}

% J0305 PARALLEL
\begin{deluxetable*}{ccccccccc}
\tablecaption{\textbf{J0305 parallel field.} Same as Table \ref{table:j0305q} for the J0305 parallel. \label{table:j0305p}}
\tablehead{\colhead{ID} & \colhead{RA} & \colhead{Dec} & \colhead{photo-$z$} & \colhead{F606W} & \colhead{F814W} & \colhead{F850LP} & \colhead{F110W} & \colhead{F140W}}
\startdata
\sidehead{Secure}
3331 & 03:05:45.0 & $-$31:51:37.0 & 6.4 $\pm$ 0.2 & 0.7 $\pm$ 0.9 & 6.0 $\pm$ 1.3 & 28.9 $\pm$ 3.0 & 26.8 $\pm$ 1.3 & 27.8 $\pm$ 1.2 \\
1540 & 03:05:43.7 & $-$31:52:09.0 & 6.2 $\pm$ 0.1 & 0.6 $\pm$ 0.9 & 2.5 $\pm$ 1.2 & 19.6 $\pm$ 2.9 & 5.7 $\pm$ 1.4 & 5.5 $\pm$ 1.3 \\
4236 & 03:05:47.1 & $-$31:51:26.2 & 6.3 $\pm$ 0.2 & $-$2.1 $\pm$ 1.0 & $<$1.4 & 23.5 $\pm$ 3.5 & 5.5 $\pm$ 1.4 & 5.8 $\pm$ 1.3 \\
371 & 03:05:40.0 & $-$31:52:14.5 & 7.0 $\pm$ 0.1 & 2.7 $\pm$ 0.8 & $-$1.0 $\pm$ 1.1 & 14.4 $\pm$ 2.8 & 15.5 $\pm$ 1.2 & 13.8 $\pm$ 1.1 \\
1873 & 03:05:45.1 & $-$31:52:24.4 & 7.0 $\pm$ 0.1 & $-$1.1 $\pm$ 0.9 & 0.1 $\pm$ 1.2 & 12.5 $\pm$ 3.0 & 14.3 $\pm$ 1.3 & 12.4 $\pm$ 1.2 \\
2948 & 03:05:48.0 & $-$31:52:48.6 & 6.5 $\pm$ 0.2 & 0.6 $\pm$ 0.9 & 1.0 $\pm$ 1.3 & 11.2 $\pm$ 3.2 & 7.2 $\pm$ 1.4 & 5.4 $\pm$ 1.5 \\
571 & 03:05:42.0 & $-$31:52:29.3 & 6.5 $\pm$ 0.3 & 0.7 $\pm$ 0.9 & 0.3 $\pm$ 1.3 & 15.7 $\pm$ 3.5 & 9.2 $\pm$ 1.3 & 5.8 $\pm$ 1.2 \\
827 & 03:05:42.5 & $-$31:52:21.7 & 8.1 $\pm$ 0.4 & $-$0.2 $\pm$ 0.6 & $-$0.5 $\pm$ 0.8 & $-$0.9 $\pm$ 2.0 & 5.4 $\pm$ 0.9 & 5.5 $\pm$ 0.9 \\
\sidehead{Marginal}
3152 & 03:05:44.0 & $-$31:51:07.1 & 8.0 $\pm$ 0.5 & $-$1.6 $\pm$ 0.8 & 0.2 $\pm$ 1.1 & $-$3.6 $\pm$ 2.6 & 5.0 $\pm$ 1.1 & 4.4 $\pm$ 1.1 \\
3286 & 03:05:48.1 & $-$31:52:33.7 & 6.5 $\pm$ 0.2 & 0.4 $\pm$ 0.8 & 4.3 $\pm$ 1.1 & 2.7 $\pm$ 2.6 & 15.4 $\pm$ 1.2 & 13.0 $\pm$ 1.1 \\
2033 & 03:05:44.4 & $-$31:52:02.0 & 6.4 $\pm$ 1.0 & 0.2 $\pm$ 0.9 & 2.1 $\pm$ 1.2 & 1.0 $\pm$ 3.0 & 7.9 $\pm$ 1.3 & 4.3 $\pm$ 1.2 \\
2044 & 03:05:46.2 & $-$31:52:43.9 & 6.3 $\pm$ 0.6 & 0.2 $\pm$ 0.8 & 1.9 $\pm$ 1.1 & 4.3 $\pm$ 2.8 & 7.0 $\pm$ 1.1 & 5.0 $\pm$ 1.0 \\
3932 & 03:05:44.0 & $-$31:50:49.4 & 7.8 $\pm$ 0.8 & 0.5 $\pm$ 0.7 & $-$0.2 $\pm$ 0.9 & $-$4.4 $\pm$ 2.3 & 4.0 $\pm$ 1.0 & 3.2 $\pm$ 0.9 \\
3506 & 03:05:47.4 & $-$31:52:07.7 & 6.7 $\pm$ 1.4 & 0.1 $\pm$ 0.8 & 0.3 $\pm$ 1.2 & 2.0 $\pm$ 2.9 & 6.2 $\pm$ 1.2 & 5.1 $\pm$ 1.1 \\
4326 & 03:05:46.0 & $-$31:51:02.9 & 6.5 $\pm$ 0.9 & 0.8 $\pm$ 0.9 & 1.8 $\pm$ 1.3 & 3.2 $\pm$ 3.2 & 10.0 $\pm$ 1.3 & 7.2 $\pm$ 1.3 \\
3806 & 03:05:47.2 & $-$31:51:48.7 & 7.7 $\pm$ 0.7 & 0.8 $\pm$ 0.9 & $-$0.5 $\pm$ 1.2 & $-$1.9 $\pm$ 3.0 & 8.3 $\pm$ 1.2 & 6.7 $\pm$ 1.2 \\
3238 & 03:05:44.7 & $-$31:51:15.9 & 7.0 $\pm$ 0.3 & 1.7 $\pm$ 0.9 & $-$0.5 $\pm$ 1.3 & 7.6 $\pm$ 3.2 & 8.9 $\pm$ 1.4 & 7.6 $\pm$ 1.3 \\
1222 & 03:05:40.6 & $-$31:51:14.0 & 7.3 $\pm$ 1.0 & $-$0.9 $\pm$ 0.8 & 0.1 $\pm$ 1.0 & 1.1 $\pm$ 2.8 & 6.1 $\pm$ 1.2 & 5.4 $\pm$ 1.1 \\
3828 & 03:05:48.0 & $-$31:52:22.3 & 6.9 $\pm$ 0.5 & $-$0.3 $\pm$ 0.7 & 0.5 $\pm$ 0.9 & 3.4 $\pm$ 2.3 & 5.7 $\pm$ 1.0 & 3.9 $\pm$ 0.9
\enddata
\end{deluxetable*}

%J2054 
\begin{deluxetable*}{ccccccccc}
\tablecaption{\textbf{J2054 QSO field.} Same as Table \ref{table:j0305q} for the J2054 QSO field.  \label{table:j2054q}}
\tablehead{\colhead{ID} & \colhead{RA} & \colhead{Dec} & \colhead{photo-$z$} & \colhead{F606W} & \colhead{F814W} & \colhead{F105W} & \colhead{F125W} & \colhead{F160W}}
\startdata
\sidehead{Secure}
4772 & 20:54:02.8 & $-$0:04:48.6 & 6.8 $\pm$ 0.6 & $-$1.9 $\pm$ 1.6 & 2.0 $\pm$ 2.2 & 17.0 $\pm$ 2.7 & 10.4 $\pm$ 1.1 & 12.0 $\pm$ 1.5 \\
1389 & 20:54:07.8 & $-$0:05:04.8 & 6.5 $\pm$ 0.2 & 2.6 $\pm$ 1.3 & 6.8 $\pm$ 1.7 & 25.7 $\pm$ 2.2 & 17.9 $\pm$ 0.9 & 20.9 $\pm$ 1.3 \\
2202 & 20:54:07.1 & $-$0:06:28.0 & 5.8 $\pm$ 0.5 & 2.8 $\pm$ 1.2 & 6.8 $\pm$ 1.6 & 23.3 $\pm$ 2.1 & 22.2 $\pm$ 0.8 & 20.8 $\pm$ 1.2 \\
1378 & 20:54:00.0 & $-$0:04:29.0 & 5.7 $\pm$ 0.4 & 2.4 $\pm$ 1.2 & 10.9 $\pm$ 1.8 & 27.3 $\pm$ 2.2 & 24.3 $\pm$ 0.8 & 20.5 $\pm$ 1.2 \\
2579 & 20:54:06.2 & $-$0:05:12.2 & 6.7 $\pm$ 0.2 & $-$0.4 $\pm$ 1.2 & 3.9 $\pm$ 1.6 & 31.5 $\pm$ 2.0 & 26.9 $\pm$ 0.8 & 25.7 $\pm$ 1.2 \\
2336 & 20:54:06.2 & $-$0:04:14.3 & 6.2 $\pm$ 0.6 & 0.9 $\pm$ 1.3 & 4.0 $\pm$ 1.7 & 24.5 $\pm$ 2.1 & 27.0 $\pm$ 0.9 & 25.9 $\pm$ 1.4 \\
3273 & 20:54:06.2 & $-$0:06:08.0 & 6.5 $\pm$ 0.0 & $-$0.7 $\pm$ 1.1 & 7.8 $\pm$ 1.7 & 17.5 $\pm$ 2.0 & 12.9 $\pm$ 0.8 & 13.3 $\pm$ 1.2 \\
2700 & 20:54:05.9 & $-$0:05:13.2 & 5.9 $\pm$ 0.5 & 1.9 $\pm$ 1.1 & 13.3 $\pm$ 1.5 & 40.8 $\pm$ 1.9 & 33.7 $\pm$ 0.7 & 35.9 $\pm$ 1.1 \\
2516 & 20:54:05.9 & -0:04:19.1 & 6.5 $\pm$ 0.2 & 0.9 $\pm$ 1.3 & 6.4 $\pm$ 1.7 & 47.8 $\pm$ 2.1 & 47.6 $\pm$ 0.8 & 48.3 $\pm$ 1.2 \\
\sidehead{Marginal}
1888 & 20:54:07.1 & $-$0:05:17.0 & 6.5 $\pm$ 0.6 & 0.6 $\pm$ 1.2 & 2.7 $\pm$ 2.0 & 12.4 $\pm$ 2.0 & 7.6 $\pm$ 0.8 & 8.8 $\pm$ 1.2 \\
4695 & 20:54:02.9 & $-$0:04:47.4 & 5.6 $\pm$ 0.5 & 1.6 $\pm$ 1.2 & 6.3 $\pm$ 1.8 & 15.0 $\pm$ 2.0 & 16.4 $\pm$ 0.8 & 16.3 $\pm$ 1.2 \\
3530 & 20:54:05.0 & $-$0:06:17.0 & 6.5 $\pm$ 1.4 & $-$0.1 $\pm$ 1.3 & 0.9 $\pm$ 1.8 & 8.5 $\pm$ 2.3 & 7.7 $\pm$ 0.9 & 4.7 $\pm$ 1.4 \\
4448 & 20:54:03.7 & $-$0:06:09.4 & 7.0 $\pm$ 0.7 & $-$0.3 $\pm$ 1.2 & 0.2 $\pm$ 1.6 & 14.4 $\pm$ 2.1 & 12.6 $\pm$ 0.8 & 14.5 $\pm$ 1.2 \\
3797 & 20:54:04.1 & $-$0:04:26.3 & 7.5 $\pm$ 0.2 & 0.4 $\pm$ 1.2 & $-$0.2 $\pm$ 1.6 & 8.1 $\pm$ 1.9 & 6.9 $\pm$ 0.8 & 7.8 $\pm$ 1.1 \\
2312 & 20:54:06.7 & $-$0:05:31.0 & 6.5 $\pm$ 0.4 & 1.4 $\pm$ 1.3 & 2.9 $\pm$ 1.7 & 12.3 $\pm$ 2.1 & 7.4 $\pm$ 0.8 & 8.4 $\pm$ 1.2 \\
1349 & 20:54:07.7 & $-$0:04:52.9 & 6.6 $\pm$ 0.5 & $-$0.3 $\pm$ 1.2 & 2.2 $\pm$ 1.7 & 12.6 $\pm$ 2.3 & 8.9 $\pm$ 0.8 & 14.0 $\pm$ 1.2 \\
1105 & 20:54:08.4 & $-$0:05:30.1 & 7.8 $\pm$ $-$0.0 & 2.4 $\pm$ 1.6 & $<$2.0 & 13.1 $\pm$ 2.6 & 13.4 $\pm$ 1.1 & 14.3 $\pm$ 1.5 \\
2783 & 20:54:05.8 & $-$0:05:00.3 & 6.4 $\pm$ 0.7 & 4.6 $\pm$ 1.8 & 1.6 $\pm$ 2.2 & 17.1 $\pm$ 2.9 & 13.9 $\pm$ 1.2 & 10.3 $\pm$ 1.6
\enddata
\end{deluxetable*}

% j2054 parallel

\begin{deluxetable*}{ccccccc}
\tablecaption{\textbf{J2054 parallel field.} Same as Table \ref{table:j0305q} for the J2054 parallel.  \label{table:j2054p}}
\tablehead{\colhead{ID} & \colhead{RA} & \colhead{Dec} & \colhead{photo-$z$} & \colhead{F814W} & \colhead{F110W} & \colhead{F140W} }
\startdata
\sidehead{Secure}
897 & 20:54:21.8 & $-$0:10:13.0 & 6.3 $\pm$ 0.3 & 12.0 $\pm$ 2.1 & 49.1 $\pm$ 0.9 & 48.5 $\pm$ 1.6 \\
\sidehead{Marginal}
4551 & 20:54:27.0 & $-$0:09:47.0 & 6.6 $\pm$ 0.8 & 2.0 $\pm$ 2.5 & 10.3 $\pm$ 0.8 & 6.9 $\pm$ 1.3 \\
3684 & 20:54:25.5 & $-$0:10:00.9 & 6.3 $\pm$ 1.4 & 2.3 $\pm$ 1.8 & 10.9 $\pm$ 0.7 & 10.4 $\pm$ 1.3 \\
1028 & 20:54:21.9 & $-$0:10:00.9 & 6.9 $\pm$ 1.0 & 1.4 $\pm$ 2.5 & 10.1 $\pm$ 1.0 & 8.0 $\pm$ 1.9 \\
643 & 20:54:20.7 & $-$0:09:01.0 & 6.2 $\pm$ 0.5 & 4.9 $\pm$ 1.8 & 27.3 $\pm$ 0.8 & 31.7 $\pm$ 1.4 \\
1876 & 20:54:23.3 & $-$0:10:15.9 & 6.7 $\pm$ 1.6 & 0.9 $\pm$ 1.6 & 14.5 $\pm$ 0.6 & 16.8 $\pm$ 1.2 \\
3257 & 20:54:24.6 & $-$0:09:08.0 & 6.6 $\pm$ 0.5 & 2.8 $\pm$ 2.1 & 10.5 $\pm$ 0.8 & 7.1 $\pm$ 1.5
\enddata
\end{deluxetable*}

% j2348 qso
\begin{deluxetable*}{cccccccccc}
\tablecaption{\textbf{J2348 QSO field.} Same as Table \ref{table:j0305q} for the J2348 QSO field.  \label{table:j2348q}}
\tablehead{\colhead{ID} & \colhead{RA} & \colhead{Dec} & \colhead{photo-$z$} & \colhead{F606W} & \colhead{F814W} & \colhead{F105W} & \colhead{F125W} & \colhead{F160W} }
\startdata
\sidehead{Secure}
853 & 23:48:31.1 & $-$30:54:04.8 & 6.6 $\pm$ 0.2 & 0.2 $\pm$ 0.8 & 3.2 $\pm$ 1.2 & 25.0 $\pm$ 1.8 & 26.9 $\pm$ 1.3 & 30.9 $\pm$ 1.5 \\
1173 & 23:48:31.2 & $-$30:53:36.2 & 6.9 $\pm$ 0.4 & $-$2.5 $\pm$ 0.8 & 2.7 $\pm$ 1.2 & 32.3 $\pm$ 2.1 & 23.4 $\pm$ 1.5 & 32.1 $\pm$ 1.8 \\
1757 & 23:48:33.7 & $-$30:54:58.1 & 6.9 $\pm$ 0.5 & 1.2 $\pm$ 0.9 & 0.8 $\pm$ 1.4 & 15.7 $\pm$ 2.2 & 7.3 $\pm$ 1.6 & 8.3 $\pm$ 1.9 \\
2210 & 23:48:33.9 & $-$30:54:21.8 & 6.8 $\pm$ 1.1 & 1.1 $\pm$ 1.4 & $-$0.3 $\pm$ 2.1 & 16.6 $\pm$ 3.3 & 15.8 $\pm$ 2.3 & 11.3 $\pm$ 2.8 \\
\sidehead{Marginal}
3775 & 23:48:37.2 & $-$30:54:51.0 & 7.2 $\pm$ 0.8 & $-$0.8 $\pm$ 1.0 & $-$1.3 $\pm$ 1.5 & 7.0 $\pm$ 2.4 & 5.9 $\pm$ 1.7 & 4.9 $\pm$ 2.0 \\
4868 & 23:48:40.2 & $-$30:53:40.8 & 7.2 $\pm$ 0.7 & 1.3 $\pm$ 1.1 & $-$1.2 $\pm$ 1.7 & 11.1 $\pm$ 2.8 & 10.8 $\pm$ 2.0 & 9.1 $\pm$ 2.3
\enddata
\end{deluxetable*}

% j2348 parallel
\begin{deluxetable*}{ccccccccc}
\tablecaption{\textbf{J2348 parallel field.} Same as Table \ref{table:j0305q} for the J2348 parallel.  \label{table:j2348p}}
\tablehead{\colhead{ID} & \colhead{RA} & \colhead{Dec} & \colhead{photo-$z$} & \colhead{F606W} & \colhead{F814W} & \colhead{F850LP} & \colhead{F110W} & \colhead{F140W}}
\startdata
\sidehead{Secure}
623 & 23:48:14.0 & $-$30:50:18.2 & 7.5 $\pm$ 0.6 & $-$0.4 $\pm$ 0.7 & 0.2 $\pm$ 1.3 & 0.9 $\pm$ 2.9 & 9.3 $\pm$ 0.8 & 7.8 $\pm$ 0.8 \\
\sidehead{Marginal}
3679 & 23:48:11 & $-$30:51:30.0 & 8.0 $\pm$ 0.4 & $-$0.5 $\pm$ 0.9 & 1.3 $\pm$ 1.6 & $-$4.2 $\pm$ 3.5 & 8.6 $\pm$ 1.1 & 8.2 $\pm$ 1.1 \\
2772 & 23:48:11.7 & $-$30:50:50.9 & 7.2 $\pm$ 1.2 & 1.2 $\pm$ 1.1 & $-$2.4 $\pm$ 1.7 & 3.3 $\pm$ 3.6 & 10.1 $\pm$ 1.0 & 11.6 $\pm$ 1.0 \\
2035 & 23:48:12.8 & $-$30:50:51.4 & 6.9 $\pm$ 0.9 & $-$1.7 $\pm$ 1.1 & 1.5 $\pm$ 1.8 & 1.0 $\pm$ 4.7 & 7.2 $\pm$ 1.1 & 4.1 $\pm$ 1.1
\enddata
\end{deluxetable*}
\clearpage

\acknowledgements
JBC and CMC thank the National Science Foundation for support through grants AST-1814034 and AST-2009577 as well as the University of Texas at Austin College of Natural Sciences for support. 
JBC acknowledges support from the Beatrice Tinsley Graduate Fellowship awarded by the University of Texas at Austin Department of Astronomy.
CMC also acknowledges support from the Research Corporation for Science Advancement from a 2019 Cottrell Scholar Award sponsored by IF/THEN, an initiative of Lyda Hill Philanthropies.

All of the \textit{HST} data presented in this paper were obtained from the Mikulski Archive for Space Telescopes (MAST) at the Space Telescope Science Institute. 
The specific observations analyzed can be accessed via \dataset[https://doi.org/10.17909/wzeq-f536]{https://doi.org/10.17909/wzeq-f536}.
STScI is operated by the Association of Universities for Research in Astronomy, Inc., under NASA contract NAS5–26555. 
Support to MAST for these data is provided by the NASA Office of Space Science via grant NAG5–7584 and by other grants and contracts.
This work was supported by STScI \textit{HST} Cycle 27 GO\#15064.

This research has benefited from the SpeX Prism Spectral Libraries, maintained by Adam Burgasser at http://www.browndwarfs.org/spexprism.
This work made use of Astropy:\footnote{http://www.astropy.org} a community-developed core Python package and an ecosystem of tools and resources for astronomy.
This work also made use of Source Extractor\footnote{https://sextractor.readthedocs.io/en/latest/Introduction.html}.

\bibliography{main}

\appendix

\section{Zeropoint uncertainty}\label{sec:zptunc}
All of the candidate LBGs are quite blue (most with $\beta<-1$), as expected since galaxies at $6<z<8$ are typically intrinsically blue owing to younger stellar populations and lower metallicity than galaxies at lower redshift \citep[e.g.,][]{Larson2022a, Finkelstein2022a}. 
Note that several candidate LBGs (especially those in our ``marginal'' sample) appear to occupy a region of the color space that implies unphysically blue ($\beta < -3$) UV continua, but the ``blue bias" is common for high-redshift candidates close to the detection limit \citep[e.g.,][]{Rogers2013a}. 
Additionally, the very blue galaxies dominate in the J0305 and J2348 QSO fields, which had significant uncertainty on the WFC3 IR zeropoints since they were calibrated using extrapolations from stars in a different field (see $\S$\ref{sec:zpt}). 
When $\beta$ was measured using the slope of the best-fit EAZY SED, it was generally in line with high-redshift expectations ($\beta > -2.5$) though many clustered around $\beta\sim-3$. 

We tested the photometry based on theoretical zeropoints (without aligning to 2MASS) as well as adjustments based on \textit{Gaia} stars (where \textit{Gaia} $g$ band is extrapolated to the IR using $BP-RP$) which both resulted in significant systematic offsets from the best fit SEDs (e.g., \hband\, always significantly fainter than expected given the \yband$-$\jband\, slope).
When performing the same alignments using the original, non-PSF-homogenized images, the same zeropoint offsets arise, so the problem does not lie with our flux conservation or PSF-matching routine.
In the end, we kept the WFC3 alignment to J2054 QSO since the observations were taken at similar times and we assume a consistent zeropoint offset between the fields (and indeed the offsets between \textit{Gaia} and ACS were consistent across the three fields).
We note that the changes we made to the zeropoints result 
in the same sources always passing our selection criteria since the modified flux measurements were within the margin of error and the SNR remains the same regardless of zeropoint, so the only change is to the slope of the SED.
Therefore, the absolute IR photometry should be taken with caution, but since we are not deriving physical properties of the galaxies from SED fitting we choose to accept the zeropoint uncertainty at face value.
Regardless, even in our robustly calibrated J2054 QSO field, all of our candidate galaxies are consistent with being quite blue, indicating young stellar populations.
Finally, we note that because the majority of our galaxies are blue but were detected using the longest-wavelength image, this could contribute to some incompleteness at the faint end of the UVLF where galaxies might scatter below the \hband\, detection limit criterion.
\section{The effect of template choice on photometric redshifts}\label{sec:templates}

One caveat to our analysis is that in the absence of spectroscopy or narrowband imaging covering Ly$\alpha$, we must make an assumption for a reasonable set of SED templates.
This sample of quasars is in the range of $6<z<7$, precisely the range of uncertainty regarding when cosmic reionization ends \citep[e.g.,][]{Robertson2021a}.
We have made the assumption that Ly$\alpha$ will be completely or partially attenuated for all of the LBGs in this sample, discarding the L22 templates which contained Ly$\alpha$ at the full strength output by \verb|Cloudy|.
A number of conditions affect the transmission of Ly$\alpha$ at these redshifts, including the neutral fraction of the IGM, the surrounding environment of a galaxy, and intrinsic properties of the galaxy facilitating the escape of Ly$\alpha$ photons, but in general, the strength of Ly$\alpha$ is predicted to be weaker at $6<z<8$ than lower-redshift LAEs at matched UV luminosity.
%; Tang23 indeed finds that LAEs at $z>7$ show a Ly$\alpha$ strength $<6\times$ that in low-redshift samples at matched UV luminosity.
On the other hand, Ly$\alpha$ transmission may be enhanced for luminous galaxies residing in overdense regions that generate enormous ionized bubbles at $z>6$ \citep[e.g.,][]{Endsley2022a, Larson2022b}.

We estimate that the template with the strongest Ly$\alpha$ emission can boost the \iband\, photometry by 0.05--0.3 mag at the same continuum level, which changes the range of available well-fitting redshifts as the Lyman break must shift to the redder end of the filter to produce a dropout at a fixed broadband flux.
To investigate the effect of template choice on the redshift solutions reported here, we ran EAZY separately for a run where we include all of the L22 templates described in $\S$\ref{sec:eazy} (including full-strength Ly$\alpha$, hereafter {\tt ALL\_TEMPS}) and one where we exclude all templates containing Ly$\alpha$ (hereafter {\tt NO\_LYA}).
Overall, the sources identified as candidate LBGs remain unchanged, but their best-fit redshifts shift slightly which changes the final list of sources considered to be part of the overdensity.
We find that the {\tt ALL\_TEMPS} run chooses a best-fit solution with no Ly$\alpha$ line in 25\% of all our LBGs, but in cases where the best-fit {\tt ALL\_TEMPS} template contains a line, the {\tt NO\_LYA} version nearly always prefers the same or a lower-redshift solution with a median shift of $\Delta z = 0.15$.
The minimum $\chi^2$s are the same or higher for the {\tt NO\_LYA} runs, but not significantly --- indeed, none of them prefer the {\tt ALL\_TEMPS} solution at $\Delta\chi^2 > 4$. 
The \pz\, distributions are generally narrower in the {\tt NO\_LYA} solutions, which shrinks the effective volume, so the value of $\delta_{\rm gal}$ rises, but since it is dominated by Poisson uncertainty the results are statistically identical within 0.5$\sigma$. 
The only shift is in $\langle z \rangle$> of the overdensity, which is $\Delta z$ = 0.25--0.5 lower in all fields, but this is within the errors of our photo-$z$'s.
We conclude that given the complete lack of constraint on Ly$\alpha$ and the relative photo-$z$ uncertainty with our current observations, the effect of template choice is mostly negligible; however, follow-up spectroscopy could reveal that the overdensities we see in J0305 and J2054 are serendipitous overdensities along the line of sight rather than associated with the quasar.
Indeed, \citet{Fujimoto2023a} selected $z>9$ LBGs from \textit{JWST}/NIRCam photometry using the same template set as {\tt ALL\_TEMPS} and found that NIRSpec followup revealed that 6/7 of these candidates had lower spectroscopic redshifts by $\Delta z = 0.25-0.5$, so this is something to keep in mind in terms of conclusions about physical associations. 
\clearpage

\end{document}